\documentclass[structabstract]{aa}  
\usepackage{graphicx}
\usepackage{color}
\usepackage[]{natbib}
\usepackage{longtable}
\usepackage{txfonts}
\begin{document}
   \title{A ring as a model of the main belt in planetary ephemerides}


   \author{P. Kuchynka\inst{1} \and  J. Laskar\inst{1} \and A. Fienga\inst{1}\fnmsep\inst{2} \and H. Manche\inst{1}}

   \institute{Astronomie et Syst\`{e}mes Dynamiques, IMCCE-CNRS UMR8028, Observatoire de Paris, UPMC, \\
              77 avenue Denfert-Rochereau, 75014 Paris, France\\
              \email{kuchynka@imcce.fr}
         \and
             Observatoire de Besan\c{c}on-CNRS UMR6213\\
             41 bis avenue de l'Observatoire, 25000 Besan\c{c}on, France 
             }

   \date{Received x; accepted x}

 
  \abstract
{}
{We assess the ability of a solid ring to model a global perturbation induced by several thousands of main-belt asteroids.}
{The ring is first studied in an analytical framework that provides an estimate of all the ring's parameters excepting mass. In the second part, numerically estimated perturbations on the Earth-Mars, Earth-Venus, and Earth-Mercury distances induced by various subsets of the main-belt population are compared with perturbations induced by a ring. To account for large uncertainties in the asteroid masses, we obtain results from Monte Carlo experiments based on asteroid masses randomly generated according to available data and the statistical asteroid model.}
{The radius of the ring is analytically estimated at 2.8 AU. A systematic comparison of the ring with subsets of the main belt shows that, after removing the 300 most perturbing asteroids, the total main-belt perturbation of the Earth-Mars distance reaches on average 246 m on the 1969 - 2010 time interval. A ring with appropriate mass is able to reduce this effect to 38 m. We show that, by removing from the main belt $\sim$240 asteroids that are not necessarily the most perturbing ones, the corresponding total perturbation reaches on average 472 m, but the ring is able to reduce it down to a few meters, thus accounting for more than 99$\%$ of the total effect.}
{}


   \maketitle

\section{Introduction}

Asteroid perturbations are considered the major obstacle for achieving a satisfactory prediction accuracy in planetary ephemerides. The asteroid problem persists despite the ephemerides being fitted to a variety of highly accurate observations available over almost 40 years. The most constraining observations are tracking data from spacecraft orbiting Mars, which over the past 8 years are available with metric precision. Despite this accuracy, today's ephemerides are unable to extrapolate the Mars ranging data one year into the future with a precision better than 15 m \citep[see][]{folkner2008,fienga2009}. The obstacle to improving the situation is to correctly account for a large number of similar asteroid effects without any accurate knowledge of the asteroid masses. In modern ephemerides, usually about 300 asteroids are modeled individually and the rest of the main-belt is represented by a circular ring. Our objective in this study is to estimate the ability of a ring to model large numbers of asteroids.

In section \ref{sec_ring} we set a simple analytical framework and show that a ring is in fact a first-order representation of the main-belt perturbation. Section \ref{sec_num} describes a new implementation of the ring in the INPOP ephemeris \citep[see][]{fienga2009}. To evaluate the ring model, we chose to compare the perturbation induced by the ring on the Earth-Mars distance with the perturbations induced by a test model containing thousands of asteroids. Section \ref{sec_mb_model} describes the chosen test model that consists of 24635 asteroids assigned each according to the available data with reasonable random mass values. In section \ref{sec_comp} the effect of the ring is compared with the perturbation induced by the test model after removing various asteroid subsets. The comparison is repeated for different asteroid masses in Monte-Carlo-like experiments. We are thus able to estimate the number of asteroids that need to be individualized in the asteroid model so the ring is able to satisfactorily represent the remaining global perturbation. We also derive estimates for the amplitude of the global perturbation, together with the corresponding mass of the ring and the amplitude of the residual perturbation unmodeled by the ring. The perturbations on the Earth-Mercury and Earth-Venus distances are considered at the end of the section. Some applications of the obtained results are discussed in section \ref{sec_appli} . 
 
\section{Analytical approach} \label{sec_ring}

\subsection{Averaging the main belt} \label{ssec_analy0}

The perturbation induced on a planet by the main belt is the sum of the perturbations induced by the asteroids within the belt. Accounting in a model for each asteroid individually involves dealing with a very large number of unknown and highly correlated parameters. In consequence, it is not possible to  implement all the individual main-belt asteroids in the model. We show that with some hypotheses the asteroid model can actually be reduced to the average effect of a single object. 

\begin{table}[t]
\caption{Parameters of an averaged orbit representing the main-belt perturbation as determined from the perturbation on the inner planets.}\label{tab_ringparam}
\centering\renewcommand{\arraystretch}{1.8}
\begin{tabular*}{\columnwidth}{@{\extracolsep{\fill}} l c c c c c c}
\hline
\hline
planet & $a$' [AU] & $e$' & $\varpi$' [$^{\circ}$] & $I$' [$^{\circ}$] & $\Omega$' [$^{\circ}$]\\
\hline
Mars  & 2.71 & 0.04 & -1.30 & 0.96 & -60.34 \\
Earth & 2.74 & 0.04 & -4.88 & 0.89 & -50.31 \\
Venus & 2.75 & 0.04 & -4.03 & 0.89 & -51.08 \\
Mercury & 2.75 & 0.04 & -5.71 & 0.88 & -50.67 \\
\hline
\hline
\end{tabular*}
\end{table}

Let us consider a main-belt asteroid of mass $m'$ on a fixed orbit perturbing a planet of mass $m$. We denote the classical orbital elements with $(a,\lambda,e,\varpi,I,\Omega)$. In this whole section we use the convention of marking the variables related to the asteroid with a prime and the variables related to the planet without a prime. We assume that in the main-belt the considered asteroid is not alone on its orbit, but many other objects with similar masses are spread uniformly in terms of mean longitude on similar orbits. It is then possible to average the perturbation of the considered asteroid over the mean longitude $\lambda'$ without losing any part of the asteroid's contribution to the main-belt perturbation. For the sake of simplicity, we also average the perturbation over the mean longitude of the planet. We are thus left only with a secular effect.


\cite{laskar1995} provide the secular Hamiltonian of the three-body problem expanded in eccentricity and inclination. In the case where the semi major axis of the asteroid is always greater than the semi major axis of the planet, the Hamiltonian can be rewritten up to the second degree as
\begin{eqnarray*}
&& F(\Lambda,\lambda,H,h,Z,\zeta) = \\
\\
&& - \frac{G m m'}{a'}  \Big(  \frac{1}{2} b_{1/2}^{(0)} + \frac{\alpha}{4} b_{3/2}^{(1)} \big( \frac{H-Z}{\Lambda} + \frac{H'-Z'}{\Lambda'} \big) \\
\\
&& + \frac{\alpha}{2} b_{3/2}^{(1)} \sqrt{\frac{Z Z'}{\Lambda \Lambda'}} \cos{(\zeta-\zeta')} \\
\\
&& + \big( \frac{3 \alpha}{2} b_{3/2}^{(0)} - (\alpha^2+1) b_{3/2}^{(1)} \big) \sqrt{\frac{H H'}{\Lambda \Lambda'}} \cos{(h-h')}\Big)
\end{eqnarray*}
where $b_{1/2}^{(0)},b_{3/2}^{(0)},b_{3/2}^{(1)}$ are Laplace coefficients depending on the semi-major axis ratio $\alpha=a/a'<1$. The canonical variables $(\Lambda,\lambda,H,h,Z,\zeta)$ are obtained by a linear transformation from the variables of Delaunay:
\begin{eqnarray*}
\Lambda &=& m \sqrt{\mu a} \\
\nonumber \\
H &=& m \sqrt{\mu a} (1-\sqrt{1-e^2}) \\
\nonumber \\
Z &=& m \sqrt{\mu a (1-e^2)} (1 - \cos{I}) \\
\nonumber \\
h &=& - \varpi  \mbox{, } \zeta = - \Omega .
\end{eqnarray*}
We can calculate the perturbation induced on the planet by the asteroid from Hamilton's equations. By only keeping the lowest order terms in eccentricity and inclination, we get
\begin{eqnarray}
\dot{a} &=& 0 \nonumber \\
\nonumber \\
\dot{\lambda} &=& n \alpha^2 \frac{m'}{m_\odot} \left( \alpha b_{3/2}^{(0)} - b_{3/2}^{(1)} \right) \nonumber \\
\nonumber \\
\dot{e} &=& n \alpha^2 \frac{m'}{4 m_\odot} e' \left( 3 b_{3/2}^{(0)} - 2 (\alpha+\frac{1}{\alpha}) b_{3/2}^{(1)} \right) \sin{(\varpi - \varpi')} \nonumber \\
\nonumber \\
\dot{\varpi} &=&  n \alpha^2 \frac{m'}{4 m_\odot} \left( b_{3/2}^{(1)} + \left(\frac{e'}{e}\right) \left( 3 b_{3/2}^{(0)} - 2 (\alpha+\frac{1}{\alpha}) b_{3/2}^{(1)} \right) \cos{(\varpi - \varpi')} \right)  \nonumber \\
\nonumber \\
\dot{I} &=& n \alpha^2 \frac{m'}{4 m_\odot}  I' b_{3/2}^{(1)} \sin{(\Omega - \Omega')} \nonumber \\
\nonumber \\
\dot{\Omega} &=& - n \alpha^2 \frac{m'}{4 m_\odot} b_{3/2}^{(1)} \left( 1 - \left(\frac{I'}{I}\right) \cos{(\Omega-\Omega')}\right).   \label{eq_analy0}
\end{eqnarray}
To obtain the perturbation induced by the entire main belt, the above equations have to be summed over all the asteroids. For $\dot{\lambda}$ we can write 
\begin{eqnarray*}
\dot{\lambda}_{tot} &=& \sum_{i=1}^{N_m} \dot{\lambda_i} = N_m \left\langle  n \alpha^2 \frac{m'}{m_\odot} \left( \alpha b_{3/2}^{(0)} - b_{3/2}^{(1)} \right) \right\rangle 
\end{eqnarray*}
where $\langle\rangle$ represent the average over all the asteroids, and $N_m$ denotes the total number of modeled objects. If we assume that there is no correlation between mass and orbital elements we obtain
\begin{eqnarray*}
\dot{\lambda}_{tot} &=& N_m \big\langle m' \big\rangle \left\langle  \frac{n \alpha^2}{m_\odot} \left( \alpha b_{3/2}^{(0)} - b_{3/2}^{(1)} \right) \right\rangle \\
\\
&=& m_{tot}' \left\langle  \frac{n \alpha^2}{m_\odot} \left( \alpha b_{3/2}^{(0)} - b_{3/2}^{(1)} \right) \right\rangle .
\end{eqnarray*}
Thus $\dot{\lambda}_{tot}$ can be obtained by averaging the $\dot{\lambda}$ term with $m'=1$  in equations (\ref{eq_analy0}) and multiplying the result by the total mass of all the main-belt asteroids denoted here by $m_{tot}'$. The orbits of the asteroids are determined fairly well. We use asteroid orbits available in the ASTORB\footnote{The asteroid orbits are calculated for the 27 October 2007.} database to calculate 
\begin{eqnarray*}
\frac{\dot{\lambda}_{tot}}{m_{tot}'} &=& \frac{1}{N_m} \sum_{i=1}^{N_m} \frac{n \alpha^2}{m_\odot} \left( \alpha b_{3/2}^{(0)} - b_{3/2}^{(1)} \right) 
\end{eqnarray*}
 numerically for all the inner planets. We consider in the catalog only those asteroids with absolute magnitude below 14 and semi-major axis below 3.5 AU. Thus the estimation is based on a total of 24634 orbits. There are actually 24635 asteroids that satisfy these criteria. We eliminated 433 Eros from the selection because its semi-major axis is lower than the semi-major axis of Mars and thus equations (\ref{eq_analy0}) do not apply. 

The method described to calculate $\dot{\lambda}_{tot}/m_{tot}'$ is applied also to the other orbital parameters. We thus obtain the numerical values of $\dot{e}_{tot}/m_{tot}'$, $\dot{\varpi}_{tot}/m_{tot}'$, $\dot{I}_{tot}/m_{tot}'$ and $\dot{\Omega}_{tot}/m_{tot}'$. For each inner planet, we can find an average equivalent orbit that will perfectly fit all the calculated total perturbations. Because $\dot{\lambda}$ is a monotonously decreasing function of $\alpha$, the value of $\dot{\lambda}_{tot}/m_{tot}'$ determines the semi-major axis of the equivalent orbit unambiguously. Once the semi-major axis is determined, the calculated values of $\dot{e}_{tot}/m_{tot}'$, $\dot{\varpi}_{tot}/m_{tot}'$ determine the eccentricity and perihelion. Similarly $\dot{I}_{tot}/m_{tot}'$, $\dot{\Omega}_{tot}/m_{tot}'$ determine the inclination and node. Table \ref{tab_ringparam} summarizes the orbital parameters of the equivalent orbit obtained on all the inner planets by averaging the 24634 asteroid orbits. The chosen reference frame is a nominal invariable plane defined by an inclination of $23^{\circ}00'32"$ and a node at $3^{\circ}51'9"$ with respect to the International Celestial Reference Frame. Equations (\ref{eq_analy0}) contain only first-order terms in eccentricities and inclinations. Because of this approximation, the parameters in Table \ref{tab_ringparam} vary from one planet to another.

The effects of an orbit averaged over its mean longitude are entirely equivalent to the effects induced by a solid ring. This equivalence has been known since the works of Gauss \citep[see for example][]{hill1882} and is valid in both the secular and non-secular cases. For practical purposes, we refer in the following more often to a ring than to the effect of an averaged orbit, but the reader has to bear in mind that both are equivalent.

Although we observe non-zero eccentricity values in Table \ref{tab_ringparam}, in this work we study the modeling of the main belt with a circular ring. The non-zero eccentricities are a consequence of the tendency of asteroid perihelia to accumulate close to the longitude of the perihelium of Jupiter. The precession rate of an asteroid perihelium is in fact at its lowest when close to the perihelium of Jupiter. The asteroid perihelia thus spend more time in this region, which explains their apparent accumulation \citep[see][]{murray1999}. For the remainder of this paper we fix the radius of the ring at 2.8 AU. This value is within 0.1 AU of all the radii in Table \ref{tab_ringparam}, and it is also the value of the radius of the ring adopted in INPOP06 \citep{fienga2008}. The initial inclination of the ring with respect to the invariable plane is chosen at zero. This fixes all the ring's geometrical parameters. The only parameter of the main-belt model that is left undetermined is the mass of the ring or equivalently the total mass of the modeled asteroids.

\subsection{Perturbation of the orbital elements} \label{ssec_analy}

We use equations (\ref{eq_analy0}) to obtain analytical expressions of the secular perturbation induced by a ring on a planet. We stress that, although the analytical results are presented in the secular case, it is merely for convenience and the objective in this study is to assess the ring model in the general non-secular case. By setting $e'$ and $I'$ at zero one obtains
\begin{eqnarray}
\dot{a} &=& \dot{e} = \dot{I} = 0 \nonumber \\
 \nonumber \\
\dot{\lambda} &=& n \alpha^2 \frac{m_{tot}'}{m_\odot} ( \alpha b_{3/2}^{(0)} - b_{3/2}^{(1)} )\nonumber \\
 \nonumber \\
\dot{\varpi} &=&  n \alpha^2 \frac{m_{tot}'}{4 m_\odot} b_{3/2}^{(1)}   \nonumber \\
 \nonumber \\
\dot{\Omega} &=& -n \alpha^2 \frac{m_{tot}'}{4 m_\odot} b_{3/2}^{(1)}  . \label{eq_analy}
\end{eqnarray}
The ring induces linear drifts in the mean, perihelion, and node longitudes of the planet. There is no secular effect on the semi-major axis because, in the secular case, the Hamiltonian is independent of $\lambda$. The presence of the ring may change the mean semi-major axis by a small constant value $\Delta a_0$. This small value would translate as a shift in the mean motion $\Delta n_0$ ($=-3 n \Delta a_0/2 a$) that should be accounted for in $\dot{\lambda}$. Equations (\ref{eq_analy}) predict no perturbation of eccentricity or inclination for the second-degree Hamiltonian. To estimate the effect on these parameters, we need to consider at least a Hamiltonian of degree 4. With \cite{laskar1995} we then obtain  
\begin{eqnarray} 
\dot{e} &=& n \alpha^2 \frac{m_{tot}'}{m_\odot} I^2 e \sin{(2 \varpi - 2 \Omega)} \left(-\frac{15}{32} \alpha b_{5/2}^{(0)} + \frac{9 + 3 \alpha^2}{16} b_{5/2}^{(1)} \right) \label{eq_analybis} \\
\nonumber\\
\dot{I} &=& - n \alpha^2 \frac{m_{tot}'}{m_\odot} I e^2 \sin{(2 \varpi - 2 \Omega)} \left(-\frac{15}{32} \alpha b_{5/2}^{(0)} + \frac{9 + 3 \alpha^2}{16} b_{5/2}^{(1)} \right) .  \nonumber
\end{eqnarray}

In Table \ref{tab_estimanaly}, we use equations (\ref{eq_analy}) and (\ref{eq_analybis}) to quantitatively estimate the effect induced on Earth and Mars by a ring with radius at 2.8 AU and a mass of $0.34 \times 10^{-10} M_\odot$. These are parameters of the asteroid model in INPOP06 where the ring represents all the main-belt asteroids except for a selection of 300 individuals. The table shows that the secular effects on the eccentricities and inclinations are clearly negligible.

\begin{table}[t]
\caption{Secular effect of a ring, with INPOP06 parameters, on the orbital elements of the Earth and Mars.}\label{tab_estimanaly}
\centering\renewcommand{\arraystretch}{1.8}
\begin{tabular*}{\columnwidth}{@{\extracolsep{\fill}} c c c l}
\hline
\hline
 & Earth & Mars & units $\times 10^{11}$\\
\hline
$\dot{\lambda}$ & -1.13 & -2.69 & rad yr$^{-1}$ \\ 
$\dot{e}$ & -0.7 $\times 10^{-5}$ & -0.7 $\times 10^{-3}$ & yr$^{-1}$\\
$\dot{\varpi}$  & 0.94 & 2.65 & rad yr$^{-1}$\\
$\dot{I}$ &  0.4 $\times 10^{-5}$ &  2.3 $\times 10^{-3}$ &  rad yr$^{-1}$\\
$\dot{\Omega}$  & -0.94 & -2.65 & rad yr$^{-1}$\\
\hline
\hline
\end{tabular*}
\end{table}

\subsection{Perturbation of the Earth-Mars distance} \label{ssec_emd}

In this work we focus on the Earth-Mars mutual distance, which today is the most accurately observed parameter. Indeed Mars has been a target for many missions. These provide highly accurate ranging data with accuracies varying from roughly 20 m for the Viking data to accuracies of about 1 m for more recent missions like Mars Global Surveyor, Mars Orbiter, or the ongoing Mars Express mission. We limit our study of the asteroid perturbations to the interval between years 1969 and 2010. This corresponds roughly to the interval spanned by the available ranging data.


Let us denote with $\Delta D$ any perturbation induced on the Earth-Mars distance. We have 
\begin{eqnarray*}
\Delta D = D - D_0 ,
\end{eqnarray*}
where $D$ and $D_0$ are the Earth-Mars distances in the perturbed and unperturbed cases respectively, and $D$ and $D_0$ each depend on the perturbed and unperturbed orbital elements of Earth and Mars. By writing a first-order Taylor expansion of $D$ in terms of the perturbations of the individual orbital elements, we obtain
\begin{eqnarray}
\Delta D &=& \frac{\partial D}{\partial a_M} \Delta a_M + \frac{\partial D}{\partial e_M} \Delta e_M \;\;\;+ \frac{\partial D}{\partial i_M} \Delta i_M  \nonumber \\
\nonumber \\
&+& \frac{\partial D}{\partial \lambda_M} \Delta \lambda_M + \frac{\partial D}{\partial \varpi_M} \Delta \varpi_M + \frac{\partial D}{\partial \Omega_M} \Delta \Omega_M \nonumber \\
\nonumber \\
&+& \frac{\partial D}{\partial a_E} \Delta a_E \;+ \frac{\partial D}{\partial e_E} \Delta e_E \;\;\;\;+ \frac{\partial D}{\partial i_E} \Delta i_E \nonumber \\
\nonumber \\
&+& \frac{\partial D}{\partial \lambda_E} \Delta \lambda_E \;+ \frac{\partial D}{\partial \varpi_E} \Delta \varpi_E \;+ \frac{\partial D}{\partial \Omega_E} \Delta \Omega_E  . \label{eq_expan}
\end{eqnarray}
The indexes E and M refer to the Earth and Mars. In the expansion, ($\Delta a$, $\Delta \lambda$..., $\Delta \Omega$) represents the perturbations of the planetary orbital elements. We have, for example, for the perturbation in mean longitude
\begin{eqnarray*}
\Delta \lambda = \lambda - \lambda_0 
\end{eqnarray*}
where $\lambda$ and $\lambda_0$ are the evolutions of the mean longitude in the perturbed and unperturbed cases, respectively. In an analogous way, we can define the perturbations of the other orbital elements. 

We calculated estimates of the partial derivatives by differentiating a second-degree eccentricity and inclination expansion of the Earth-Mars distance. The resulting estimates are functions of approximately constant amplitude oscillating with frequencies close to the Earth-Mars synodic frequency. Table \ref{tab_diff} lists the amplitudes of all the partial derivatives. The significant differences among the amplitudes of the partial derivatives stem from the various orders in eccentricity and inclination. 

The general expression (\ref{eq_expan}) can be used to estimate the effect induced on the Earth-Mars distance by the ring with INPOP06 parameters. With Table \ref{tab_diff} and the values of the secular perturbations calculated in Table \ref{tab_estimanaly}, we find that the perturbation will reach approximately 5 m over one year, mainly because of the drift in the mean longitude of Mars.
 
\begin{table}[t]
\caption{The Earth-Mars distance dependence on the orbital elements.}\label{tab_diff}
\centering\renewcommand{\arraystretch}{1.8}
\begin{tabular*}{\columnwidth}{@{\extracolsep{\fill}} c c c l}
\hline
\hline
& Earth & Mars & units $\times 10^{-11}$\\
\hline
max $\left\vert \frac{\partial D}{\partial a} \right\vert$ & 1.52 & 1.63 & m AU$^{-1}$ \\ 
max $\left\vert \frac{\partial D}{\partial \lambda} \right\vert$ & 1.52 & 1.87 & m rad$^{-1}$\\
max $\left\vert \frac{\partial D}{\partial e} \right\vert$ & 3.03 & 3.56 &  m \\
max $\left\vert \frac{\partial D}{\partial \varpi} \right\vert$ & 0.05 & 0.35 & m rad$^{-1}$\\
max $\left\vert \frac{\partial D}{\partial i} \right\vert$ & 0.15 & 0.14 & m rad$^{-1}$\\
max $\left\vert \frac{\partial D}{\partial \Omega} \right\vert$ & 0.00 & 0.01 & m rad$^{-1}$\\
\hline
\hline
\end{tabular*}
\end{table}

\section{Numerical integrations} \label{sec_num}

\subsection{The implementation of the ring in INPOP}

In \cite{krasinsky2002} the ring was numerically implemented as a perturbing acceleration in the ecliptic plane of each planet, and this implementation has been adopted in INPOP06 as well. \cite{fienga2009} show that in long-term integrations the ring causes a slight drift in the Solar System barycenter of approximately 10 m over a century. Indeed with the ring modeled as an exterior force, the system's total linear momentum is not conserved. The existence of this drift motivated a more realistic implementation of the ring adopted since in INPOP08. 

The ring is treated as a solid rotating body, which fully interacts with the planets. Although its initial orientation is taken to be parallel to the system's invariable plane, the orientation of the ring is an integrated parameter that evolves with time under the influence of the planetary perturbations. The angular momentum of the ring is constant in amplitude and determined by the radius of the ring and Kepler's law. The linear and angular momenta of the system are thus conserved, which eliminates the barycenter drift occurring in the previous implementation. Because the averaged orbit is bound to the Sun by gravitation, in INPOP we fix the center of the ring to the barycenter of the Sun. We note that a free floating ring centered on the Sun or on the barycenter of the Solar System is actually unstable and would gradually drift away from its initial position. The expression for the force exerted by a ring on a particle is given in Appendix \ref{app_force}. 

\subsection{Comparison with the analytical expressions}

\begin{figure}
\centering
\includegraphics[width=9cm]{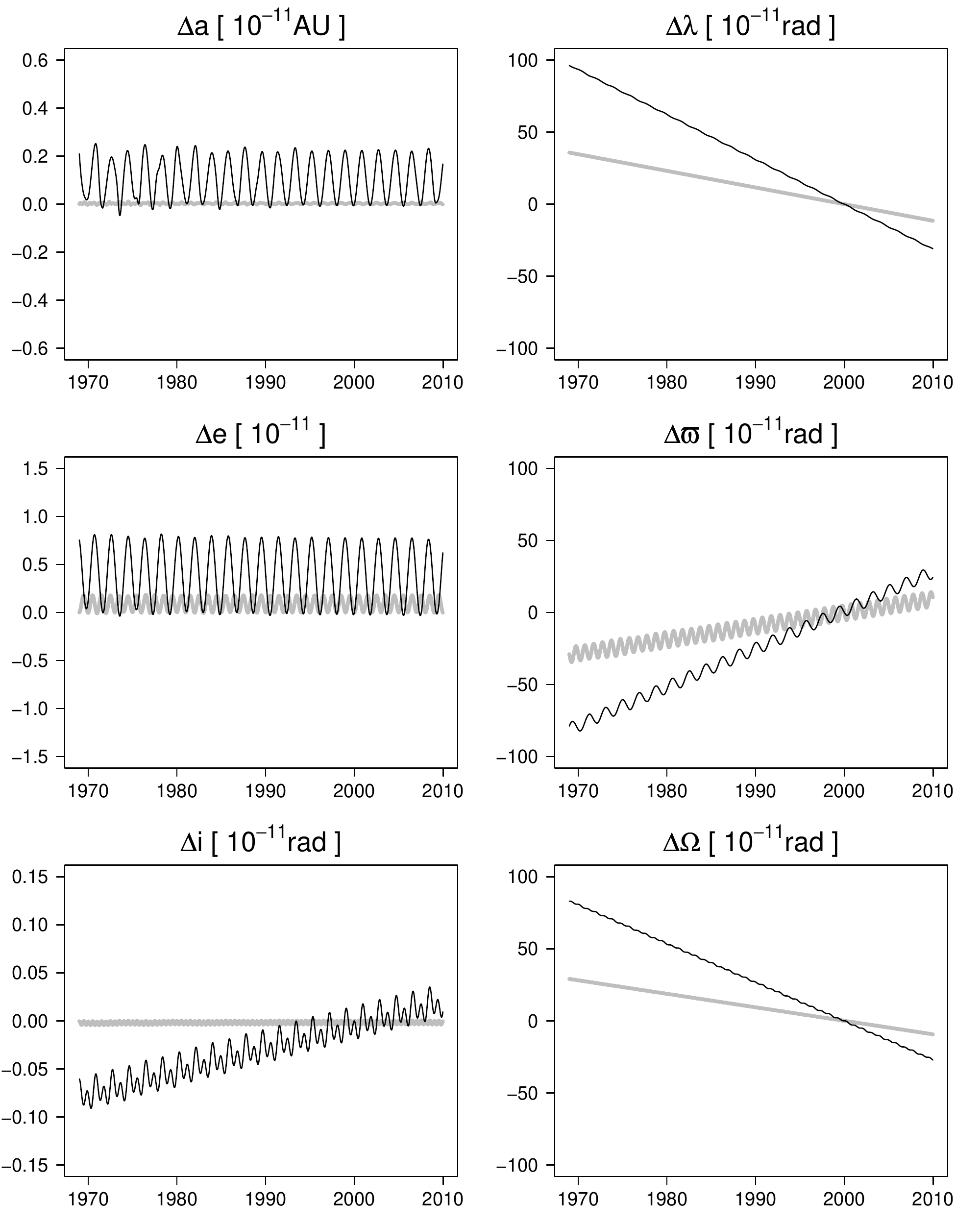}
\caption{Perturbation induced on the orbital elements of Earth (in gray) and Mars by a ring with INPOP06 parameters. }
\label{fig_Dx_ring}
\end{figure}

We used INPOP to numerically estimate the effects of a ring with INPOP06 parameters on the Earth and Mars. These effects can be isolated from the perturbations induced by other objects in the INPOP model by comparing a Solar System integration with the ring and a reference integration in which the ring is absent. The difference between the evolutions of orbital elements in both integrations provides the perturbation induced exclusively by the ring. Results obtained with this method on the 1969-2010 interval are shown in figure \ref{fig_Dx_ring}. Because integrations both with and without the ring start from identical initial conditions at J2000, all the perturbations are at zero for this date. A linear regression provides values for the numerically observed secular drifts : 
\begin{eqnarray*}
\dot{\lambda}_E &=& -1.15 \times \mbox{$10^{-11}$ rad yr$^{-1}$ }\;,\;\;\; \dot{\lambda}_M = -3.11 \times \mbox{$10^{-11}$ rad yr$^{-1}$}\\
\dot{\varpi}_E &=& 0.95 \times \mbox{$10^{-11}$ rad yr$^{-1}$ }\;\;\;,\;\;\; \dot{\varpi}_M = 2.65 \times \mbox{$10^{-11}$ rad yr$^{-1}$} \\
\dot{\Omega}_E &=& -0.94 \times \mbox{$10^{-11}$ rad yr$^{-1}$ }\;,\;\;\; \dot{\Omega}_M = -2.69 \times \mbox{$10^{-11}$ rad yr$^{-1}$} .
\end{eqnarray*}
With the exception of the mean longitude of Mars, the observed drifts agree well with the analytical predictions of Table \ref{tab_estimanaly}. For Mars, the discordance with the predicted drift in mean longitude is due to the non zero mean value of the perturbation of the semi-major axis of Mars. A constant shift in the semi-major axis corresponds to a shift in mean motion $\Delta n_0$. We can estimate $\Delta n_0$ directly from the mean value of the semi-major axis in figure \ref{fig_Dx_ring}. Adding the estimated $\Delta n_0$ to the drift in the mean longitude of Mars calculated in Table \ref{tab_estimanaly} leads to a corrected estimation $\dot{\lambda}_M = -3.05 \times 10^{-11}$ rad yr$^{-1}$. The numerically observed drift in the inclination of Mars is $\dot{I}_M = 0.0024 \times 10^{-11}$ rad yr$^{-1}$. The interval 1969-2010 is too short to correctly estimate the secular drift in the inclination of Earth. Similarly, for both planets it is impossible to estimate the secular drifts in eccentricities. 

\begin{figure}
\centering
\includegraphics[width=9cm]{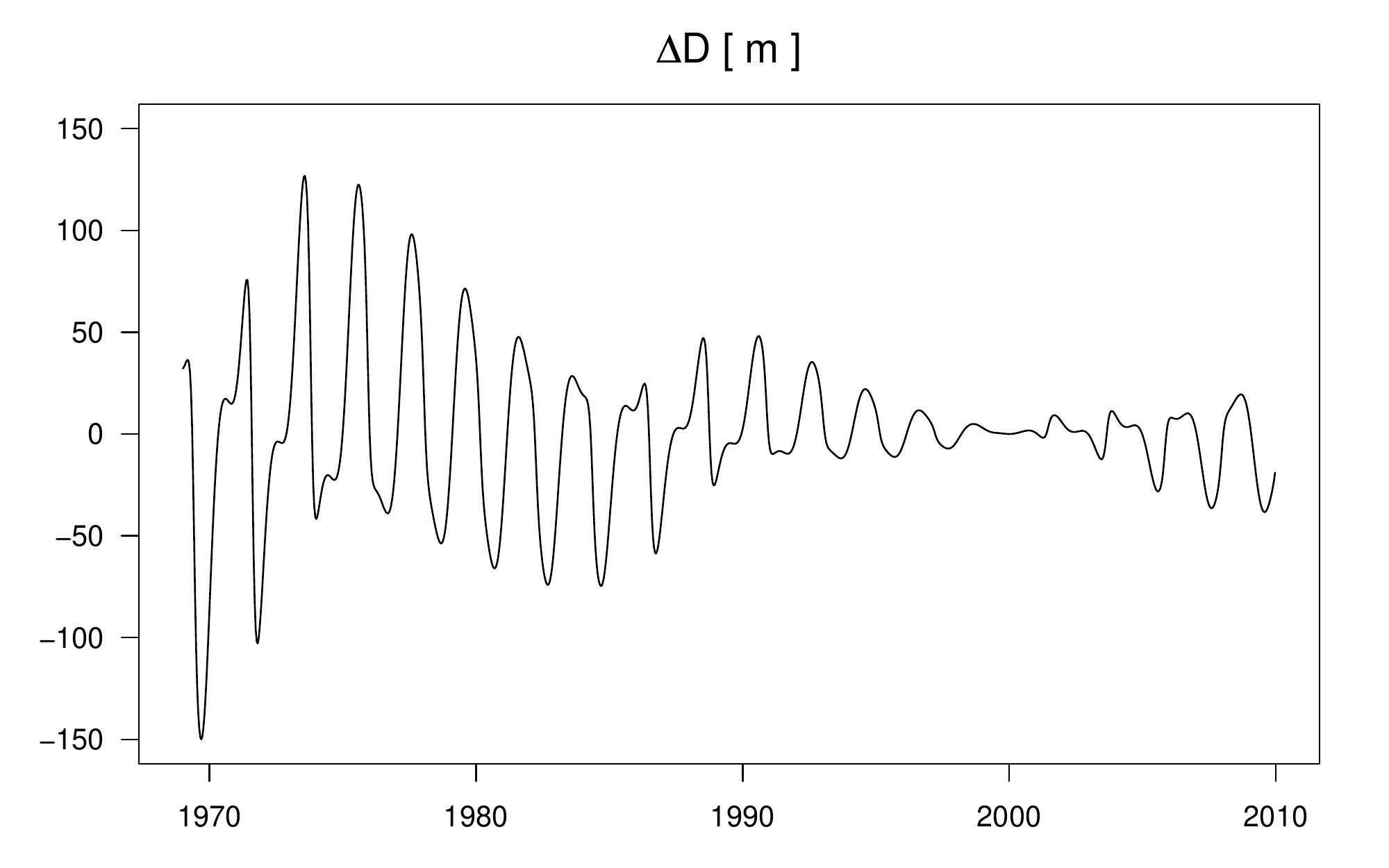}
\caption{Perturbation of the Earth-Mars distance induced by a ring with INPOP06 parameters.}
\label{fig_DD_ring}
\end{figure}

We estimate the perturbation induced by the ring on the Earth-Mars distance by comparing two integrations with and without the ring. Figure \ref{fig_DD_ring} shows that with INPOP06 parameters the effect of the ring reaches approximately 150 m. This perturbation results from secular drifts in the various orbital elements, thus the 150 m reached over 31 years between 1969 and 2000 can be translated as approximately 5 m per year. This is in good agreement with the prediction made at the end of section \ref{ssec_emd}.

\section{A test model of the main belt} \label{sec_mb_model}

From the analytical standpoint the ring appears as a first-order representation of the main belt. Nevertheless, it is rather difficult to evaluate its effective capacity to model a multitude of objects. One possible solution is to test the ring against a test model containing large numbers of individual asteroids assigned with particular masses. The ring can be tested against many different asteroid models each built with a different set of masses. Such Monte Carlo (MC) experiments can provide an estimate of the number of asteroids that have to be modeled individually in ephemerides. If the asteroid masses are individually inaccurate but globally reasonable, the MC experiments can also estimate the ring's mass.

\subsection{Asteroid selection}

We select for our test model all the asteroids in the ASTORB catalog that have an absolute magnitude brighter than 14 and are situated in terms of semi-major axis within 3.5 AU. These criteria are the same as in section \ref{ssec_analy0}. The absolute magnitude limit of 14 corresponds to the estimated completeness limit of the main-belt and NEO populations as reported by \cite{jedicke2002}. The value also leads to a relatively large but still reasonable number of objects to work with, 24635 asteroids in total. Absolute magnitude can be converted to diameter with the following formula \citep{bowell1989}:
\begin{eqnarray*}
D_{(km)} = \frac{1329}{\sqrt{\rho}} 10 ^{-0.2 H} 
\end{eqnarray*}
where D is the diameter in kilometers, H the absolute magnitude, and $\rho$ the albedo of the asteroid. With a value for a minimum possible albedo within the asteroid population, the formula sets an upper bound on the diameter of asteroids that are not included in our model. \cite{tedesco2005} estimate the mean albedo of the lowest albedo class at approximately 0.05. With this estimate, the maximum diameter of an asteroid not included in our test model is 10 km. 

\subsection{Asteroid masses} \label{ssec_astsmasses}


\begin{table}[t]
\begin{minipage}[t]{\columnwidth}
\caption{Albedo and corresponding uncertainty for objects with known taxonomy or belonging to a dynamical family.} \label{tab_tedesco}
\centering\renewcommand{\arraystretch}{1.15}\renewcommand{\footnoterule}{}
\begin{tabular*}{\columnwidth}{@{\extracolsep{\fill}} l c c }
\hline
\hline
\multicolumn{3}{c}{Family data \footnote{Values reproduced from \cite{tedesco2005}, Tables 1 and 7.}} \\
Family  & $\rho$ & $\Delta \rho$ \\
\hline
Adeona & 0.0734 & 0.0205 \\
Dora & 0.0603 & 0.0160  \\
Eos & 0.1359 & 0.0426   \\
Erigone & 0.0569 & 0.0123  \\
Eunomia & 0.1494 & 0.0864 \\
Flora & 0.2113 & 0.0905 \\
Gefion & 0.0824 & 0.0738 \\
Hygiea & 0.0515 & 0.0141  \\
Koronis & 0.2094 & 0.0603   \\
Maria & 0.2224 & 0.0525   \\
Massalia & 0.2096 & 0.0603  \\
Merxia & 0.2207 & 0.0603  \\
Themis & 0.0834 & 0.0338   \\
Veritas & 0.0693 & 0.0150   \\
Vesta & 0.2870 & 0.0795  \\
\hline
\hline
\multicolumn{3}{c}{Taxonomy data $^a$} \\
Class & $\rho$ & $\Delta \rho$ \\
\hline
Low albedo (C,G,B,F,P,T,D) & 0.0545 & 0.0345 \\
Intermediate albedo (M) & 0.1005 & 0.0155\\
Moderate albedo (S,Q) & 0.2335 & 0.1215 \\
High albedo (E,V,R) & 0.4305 & 0.0955 \\
\hline
\hline
\end{tabular*}

\end{minipage}
\end{table}

Today the total number of accurately measured asteroid masses amounts to only a few tens. Besides the case of binary objects, asteroid mass determinations are in general susceptible to systematic errors, which are hard to estimate. To assign all the selected asteroids with at least reasonable masses, we devise a simple algorithm inspired by the statistical asteroid model \citep{tedesco2005}. The algorithm processes family membership, taxonomy, and SIMPS survey data\footnote[1]{The databases are maintained in NASA's PDS Asteroid Archive.} and assigns accordingly each asteroid with an albedo $\rho$ and corresponding uncertainty $\Delta \rho$. Among the 24635 asteroids, more than two thousand asteroids have values of $\rho$ and $\Delta \rho$ directly available from SIMPS, objects with family or taxonomy data have their $\rho$ and $\Delta \rho$ attributed with data obtained in the statistical asteroid model and reproduced in Table \ref{tab_tedesco}. In cases where information overlaps, SIMPS is preferred to taxonomies and families. Taxonomy data is preferred to families, which helps to eliminate interlopers in the family data. In general at least some information is available for 10\% of the selected objects. For the remaining majority of asteroids, they are randomly assigned an albedo class with probabilities adopted from \cite{tedesco2005} (56\% low albedo, 7\% intermediate albedo, 34\% moderate albedo, and 3\% high albedo). Table \ref{tab_tedesco} defines $\rho$ and $\Delta \rho$ for each albedo class.

Asteroid diameters are calculated from corresponding $\rho$ to which a random error within $\pm \Delta \rho$ is added. For SIMPS data, the algorithm ignores the formal $\Delta \rho$ and applies instead a more realistic 10\% uncertainty directly on the diameter. As potential systematic errors in absolute magnitudes are reported by various authors \citep{juric2002}, we account for them with a $\pm 0.5$ random uncertainty added to absolute magnitude. With the lower part of Table \ref{tab_tedesco}, we can use each asteroid's mean albedo $\rho$ to assign the asteroid with a density class : C (low albedo), S (moderate albedo), and M (intermediate and high albedos merged). We stress that these density classes are attributed according to albedo. Therefore, for some objects, the density classes may not coincide with taxonomy data. Bulk porosity is expected to vary among the asteroids \citep[see][]{britt2002}, so we adopt the following intervals for the class densities : $[0.5,2.5]$ (C), $[1.6,3.8]$ (S) and $[1,5]$ (M). The density of an asteroid is chosen randomly within the corresponding class interval and together with the previously calculated diameter provides a mass. The masses of six asteroids are kept constant and equal to their published values. We fix $4.756 \times 10^{-10} M_{\odot}$, $1.025 \times 10^{-10} M_{\odot}$, $1.348 \times 10^{-10} M_{\odot}$ for 1 Ceres, 2 Pallas, 4 Vesta \citep{fienga2008}, $0.45 \times 10^{-10} M_{\odot}$ for 10 Hygiea \citep{chesley2005}, $0.03 \times 10^{-10} M_{\odot}$ for 22 Kalliope \citep{merline1999}, and $0.037 \times 10^{-10} M_{\odot}$ for 45 Eugenia \citep{margot2003}. 

We used the algorithm to generate for each asteroid a set of 100 random masses. A standard mass set is generated without any random choices. In this standard mass set, $\Delta \rho$ is put to zero and densities are maintained at INPOP06 values : 1.56 (C), 2.18 (S), and 4.26 (M). Objects without any available data are automatically considered as belonging to the C taxonomy class in the standard mass set.

\subsection{Individual perturbations and the global effect} \label{ssec_astspertu}

As in section \ref{sec_num}, the perturbation of the Earth-Mars distance is denoted with $\Delta D$. To evaluate the perturbations $\Delta D_i$ ($1 \leq i \leq 24635$) induced by each individual asteroid of the test model, we performed extensive integrations with INPOP. Each $\Delta D_i$ was obtained by comparing on the 1969-2010 interval a Solar System integration with the particular asteroid and a reference integration in which the asteroid is absent.

For a given set of asteroid masses, we can rank the asteroids according to the decreasing amplitude of their individual perturbations. Here and in the following, the amplitude of a perturbation is estimated by the maximum of $\vert \Delta D \vert$ reached on the 1969 - 2010 interval. Each $\Delta D_i$ is proportional to the mass $M_i$ of the perturbing asteroid, in consequence, we have 
\begin{eqnarray}
\Delta D_i = \frac{\partial \Delta D_i}{\partial M_i} M_i .  \label{eq_prop}
\end{eqnarray}
Equation (\ref{eq_prop}) can be rewritten as
\begin{eqnarray*}
\Delta D_i = \frac{\partial D}{\partial M_i} M_i
\end{eqnarray*}
where D is the Earth-Mars distance. This is the analog of equation (\ref{eq_expan}), but instead of considering the perturbation of the Earth-Mars distance as depending on the perturbations of the planetary orbits, we consider it as depending on the mass of the perturbing asteroid. The first-order Taylor expansion of D in terms of all the asteroid masses leads to
\begin{eqnarray*}
D(M_1, ..., M_{24635}) = D(0, ..., 0) + \frac{\partial D}{\partial M_1} M_1 + ... + \frac{\partial D}{\partial M_{24635}} M_{24635} .
\end{eqnarray*}
We can thus approximate to first-order the perturbation caused by a particular set of asteroid masses by the sum of the already calculated individual asteroid contributions:
\begin{eqnarray}
\Delta D = \Delta D_1 + ... + \Delta D_{24635} . \label{eq_scheme}
\end{eqnarray}
Although initially the $\Delta D_i$ are obtained with INPOP for the standard mass set defined in section \ref{ssec_astsmasses}, the proportionality relation (\ref{eq_prop}) provides the individual perturbations for any set of asteroid masses. In the next section we will use the expansion (\ref{eq_scheme}) to study how the resemblance between the ring, and the test model evolves with the number of asteroids removed from the test model and between different mass sets. Indeed, INPOP integrations of the Solar System with thousands of asteroids take several hours, and it is impossible to explore the parameter space by reintegrating the test model each time with new parameters. We refer to the perturbation induced by the test model after removing a selection of individuals as to the global perturbation.

\begin{figure}
\centering
\includegraphics[width=9cm]{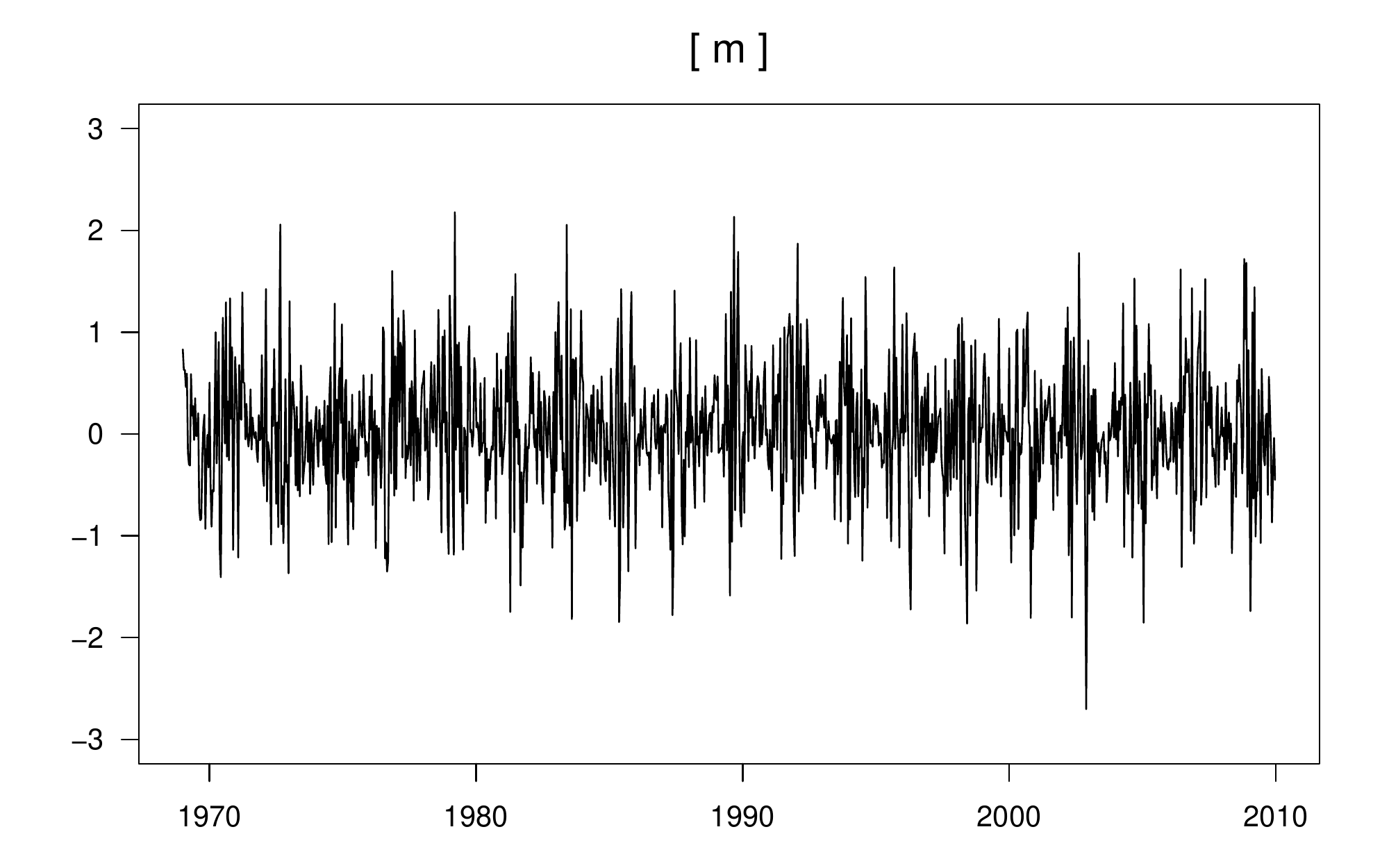}
\caption{Difference between the $\Delta D$ obtained from expansion (\ref{eq_scheme}) and the $\Delta D$ obtained from a simultaneous INPOP integration of the asteroids.}
\label{fig_err}
\end{figure}

To test the development (\ref{eq_scheme}), we can compare the global perturbation obtained from a simultaneous INPOP integration of all the asteroids, but the N most perturbing ones with the same perturbation obtained from the development. We choose N at 300 here because it corresponds to the number of asteroids usually considered individually in modern ephemerides. Figure \ref{fig_err} shows the difference between the two perturbations for the standard mass set. This difference is on the order of 1 m, which is less than 1\% of the perturbation's amplitude. In consequence, we consider the development (\ref{eq_scheme}) as satisfactory. It should be noted, that in the simultaneous integration, the mutual perturbations of the asteroids were not taken into account. 

\section{Testing the capacity of the ring to model large numbers of asteroids} \label{sec_comp}

\subsection{Selection based on amplitude} \label{sec_comp1}

\begin{figure}
\centering
\includegraphics[width=9cm]{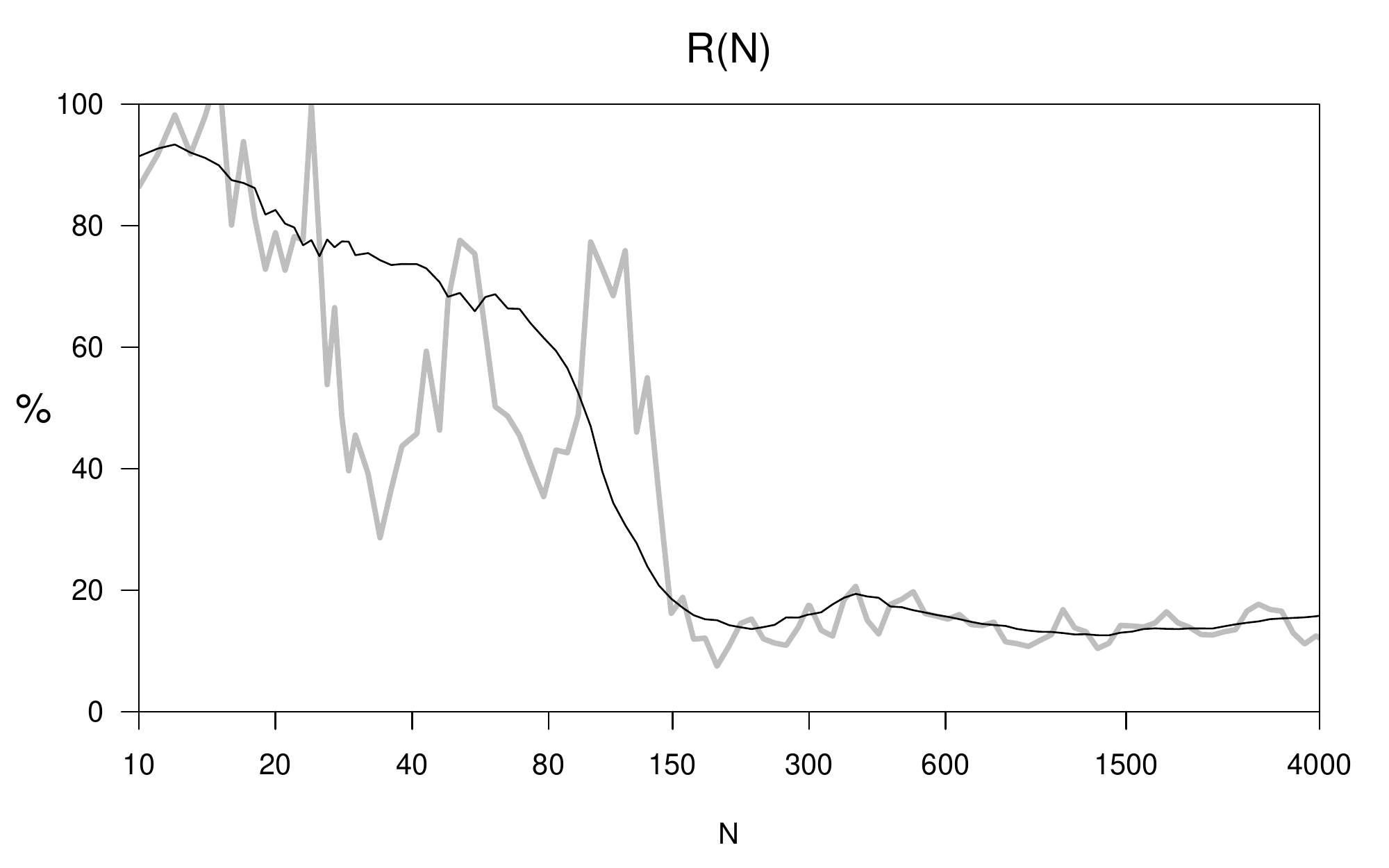}
\caption{Evolution of R(N) for the standard set of masses (in gray) and an average over 100 different sets.}
\label{fig_glob1}
\end{figure}

We denote with $\Delta D_{\mbox{\tiny{glob}}}(N)$ the global perturbation induced on the Earth-Mars distance by the test model after removing from the test model the N most perturbing asteroids. Similarly we denote with $\Delta D_{\mbox{\tiny{ring}}}$ the perturbation induced by a ring. To evaluate the capacity of the ring to represent the global perturbation, we fit for different values of N the amplitude of $\Delta D_{\mbox{\tiny{ring}}}$ so as to minimize
\begin{eqnarray*}
\left\vert \Delta D_{\mbox{\tiny{ring}}} - \Delta D_{\mbox{\tiny{glob}}}(N) \right\vert .
\end{eqnarray*}
Because $\Delta D_{\mbox{\tiny{ring}}}$ is proportional to the mass of the ring, fitting the amplitude is actually equivalent to fitting the mass of the ring. Let us denote by $R(N)$ the percentage of the amplitude of the global perturbation left after fitting the ring. We thus have
\begin{eqnarray*}
R(N) = \frac{\max \left\vert \Delta D_{\mbox{\tiny{ring}}} - \Delta D_{\mbox{\tiny{glob}}}(N) \right\vert}{\max \left\vert \Delta D_{\mbox{\tiny{glob}}} .(N) \right\vert}
\end{eqnarray*}
Figure \ref{fig_glob1} shows the evolution of $R(N)$ for the standard set of masses, as well as an average of $R(N)$ over the 100 different mass sets defined in section \ref{ssec_astsmasses}. For values of N greater than 200, the ability of the ring to model a global effect has reached its maximum and remains constant. At its best, the ring is thus able to represent more than 80$\%$ of the total perturbation amplitude. 

\begin{figure}
\centering
\includegraphics[width=9cm]{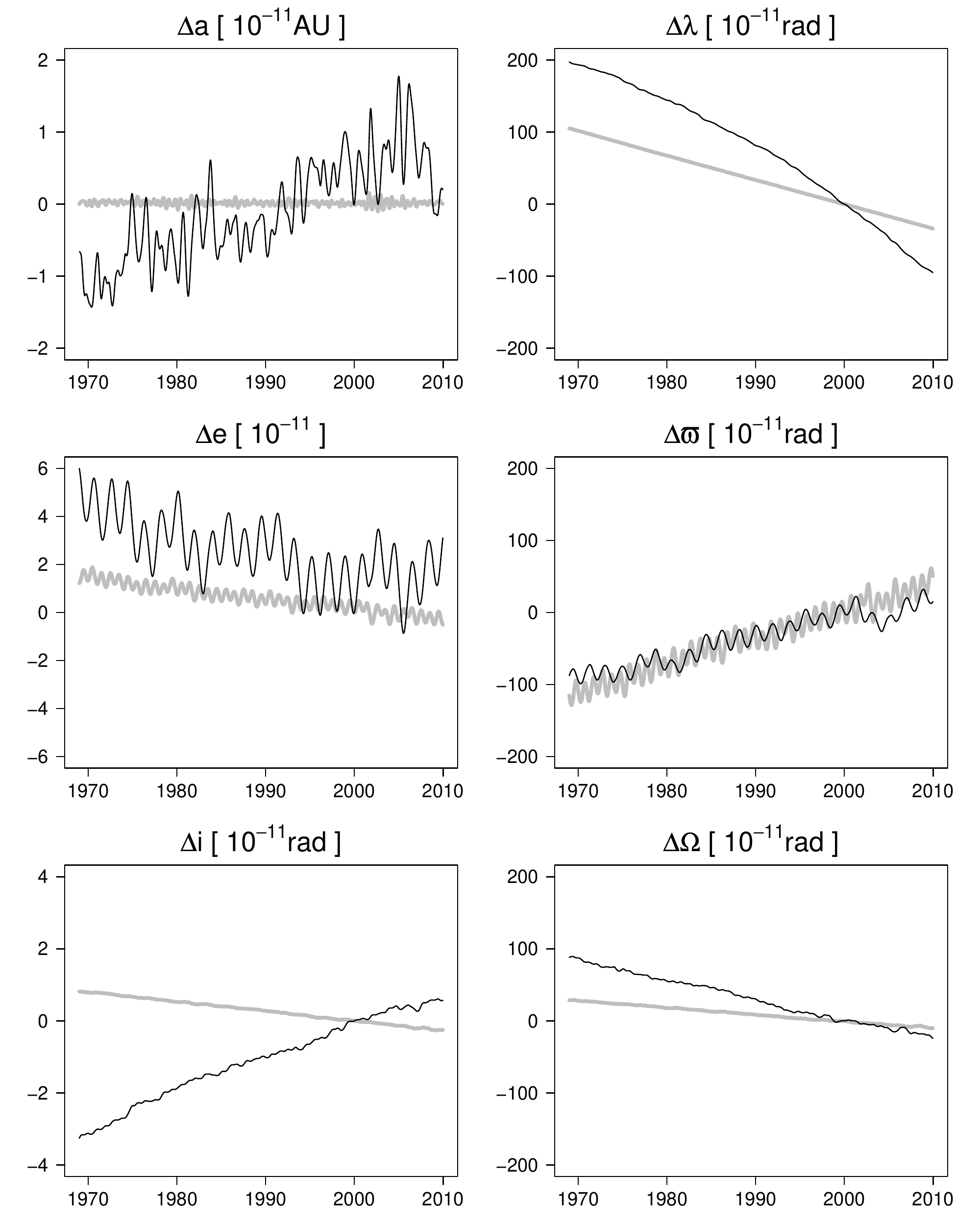}
\caption{Perturbation induced on the orbital elements of Earth (in gray) and Mars by the test model after removing from the test model the 300 most important perturbers.}
\label{fig_Dx_glob}
\end{figure}

\begin{figure}
\centering
\includegraphics[width=9cm]{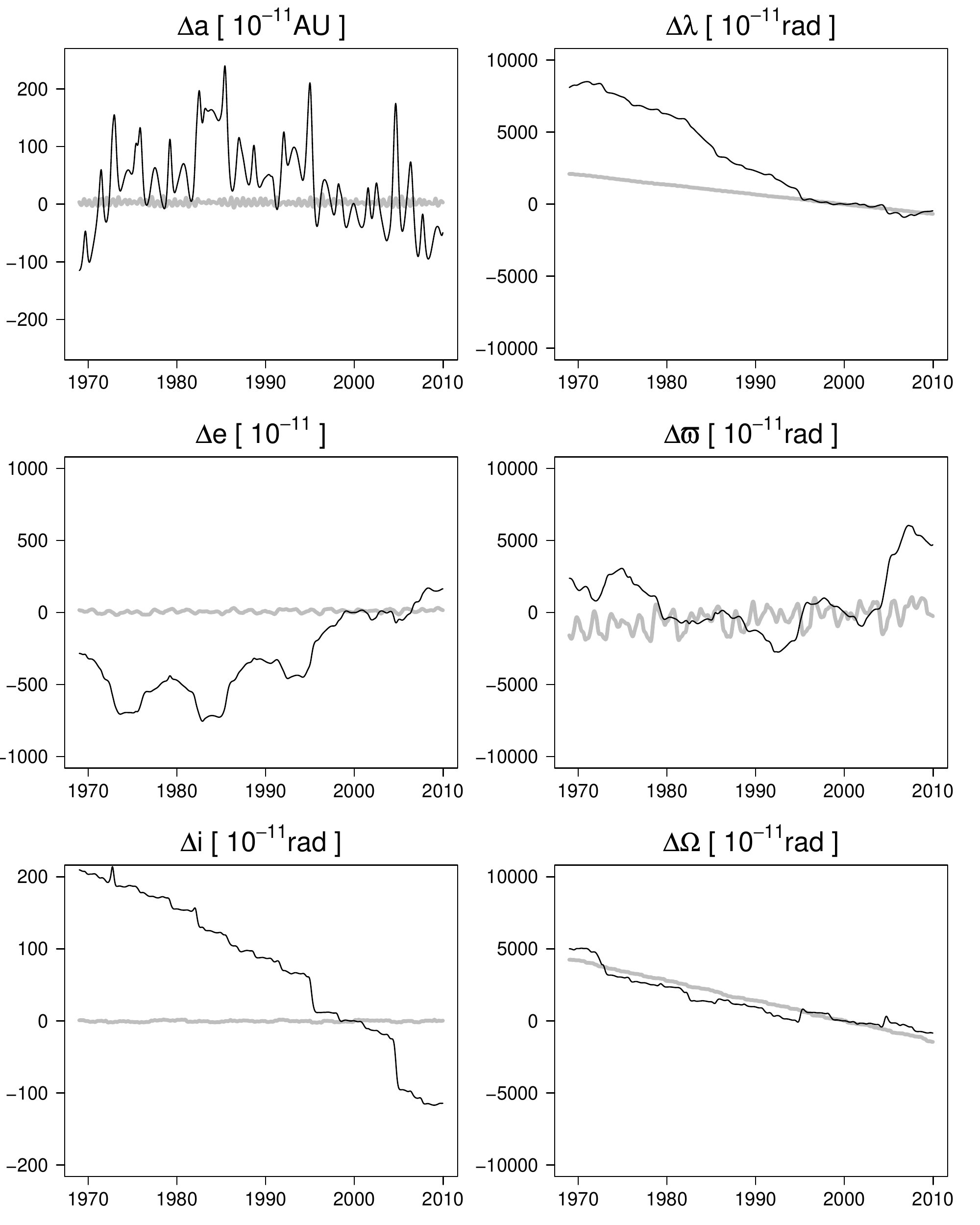}
\caption{Perturbation induced on the orbital elements of Earth (in gray) and Mars by the entire test model.}
\label{fig_Dx_glob0}
\end{figure}

We show in figure \ref{fig_Dx_glob} the effect induced on the planetary orbital elements by the test model after removing from the test model the 300 most perturbing asteroids (for the standard mass set). The observed effects are indeed rather smooth and similar to the drifts induced by a ring. For comparison, figure \ref{fig_Dx_glob0} shows the effect induced on the planetary orbital elements by all the asteroids of the test model.

\begin{figure}
\centering
\includegraphics[width=9cm]{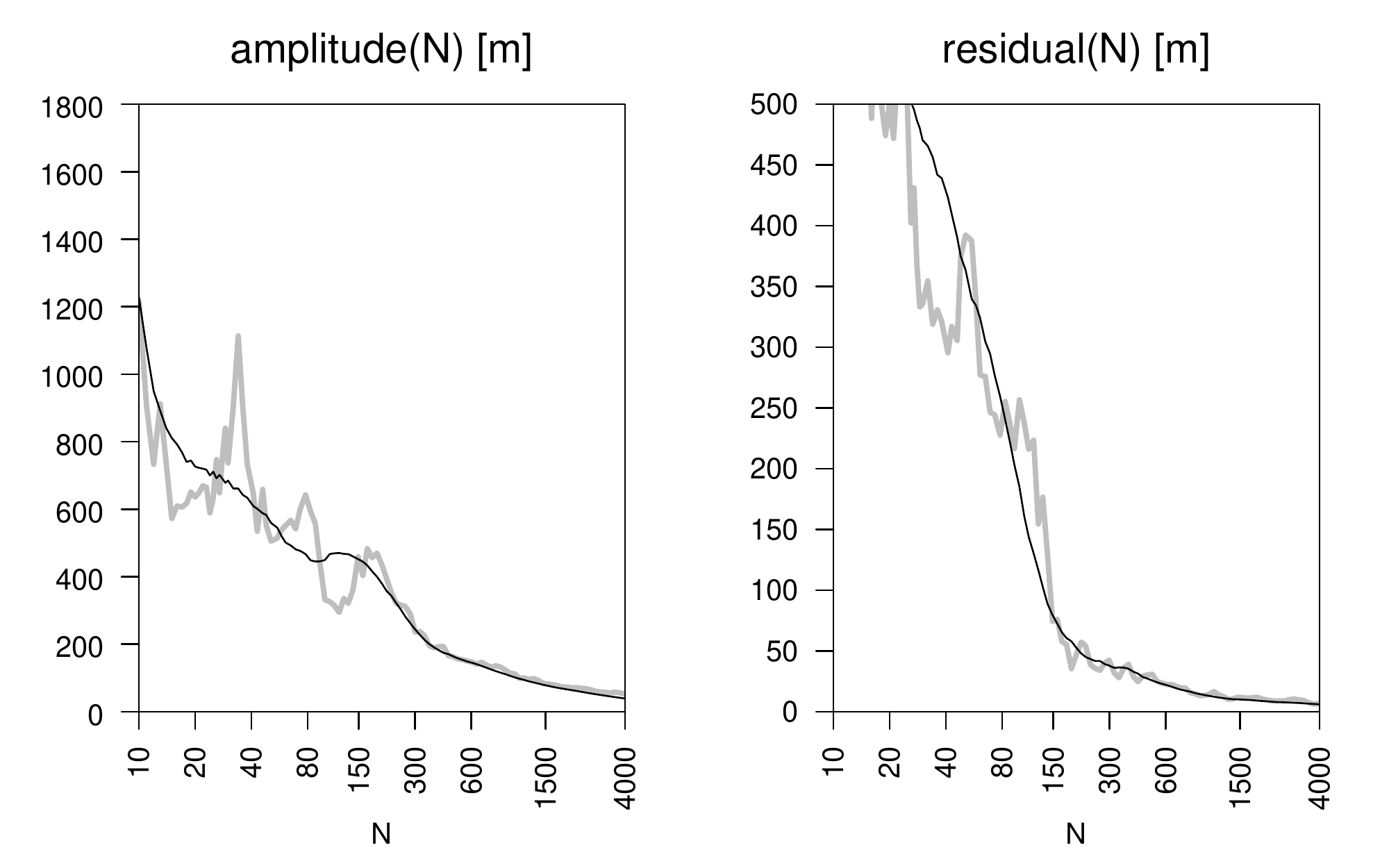}
\caption{Evolution of the amplitude of $\Delta D_{\mbox{\tiny{glob}}}$ and the amplitude of the residuals after fitting a ring for the standard set of masses (in gray) and corresponding averages computed over 100 random mass sets.}
\label{fig_ampli_res1}
\end{figure}

It is possible to calculate the average mass of the ring after removing the 300 most perturbing asteroids by minimizing 
\begin{eqnarray*}
\vert \Delta D_{\mbox{\tiny{ring}}} - \Delta D_{\mbox{\tiny{glob}}}(300) \vert .
\end{eqnarray*}
With the 100 different sets of asteroid masses, the mass of the ring is estimated at $M_{\mbox{\tiny{ring}}} = 0.6 \pm 0.2\times 10^{-10} M_{\odot}$.  In a similar way, we can estimate the average maximum reached by $\Delta D_{\mbox{\tiny{glob}}}$ at 246 m and the average maximum reached by the residuals after fitting the ring at 38 m. The obtained value of $M_{\mbox{\tiny{ring}}}$ is approximately twice as high as the mass of the ring fitted in INPOP06. Contrary to INPOP06 where parameters are fitted to observations, here the value of $M_{\mbox{\tiny{ring}}}$ is obtained solely from the test model. Figure \ref{fig_ampli_res1} shows the variations with N of the amplitude of the global perturbation and the amplitude of the residuals. The evolution of the mass of the ring is not shown because it follows proportionally the amplitude of the global perturbation.

The calculated residuals for $N = 300$ correspond to approximately 10 m over the 2000-2010 time interval. This is an order of magnitude above the residuals obtained today for the most accurate Mars ranging data. The limiting factor of the ring model is the inability to reproduce the quadratic evolution in the mean longitude of Mars (see figure \ref{fig_Dx_glob}). This quadratic evolution is in fact a consequence of the linear drift of the semi-major axis of Mars that persists in the test model for all values of N.

\subsection{The selection as a mixed integer quadratic problem} \label{sec_cplex}

\begin{figure}
\centering
\includegraphics[width=9cm]{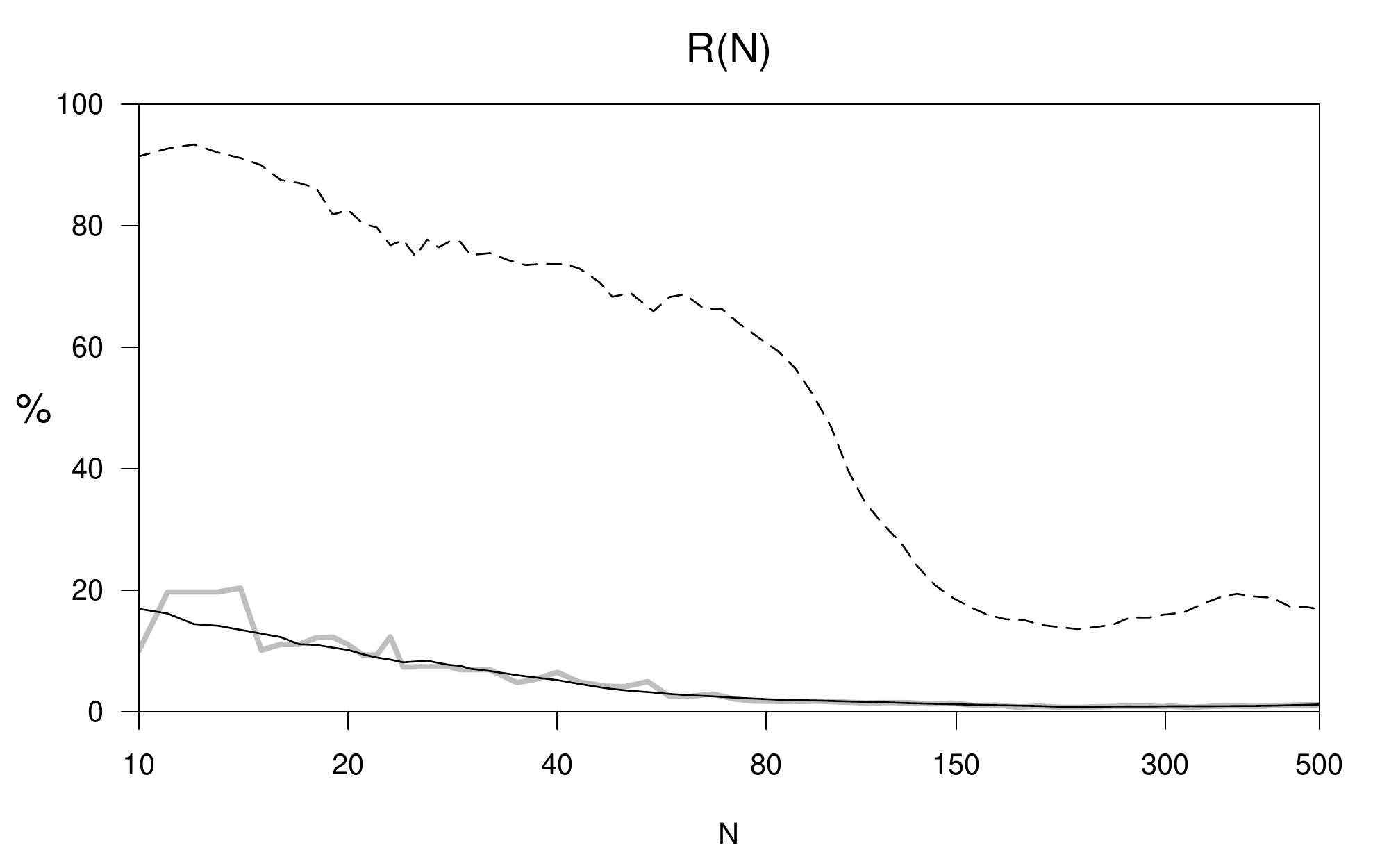}
\caption{Evolution of R(N) with the MIQP selection for the standard set of masses (in gray) and an average over 100 different sets. The dashed line represents the average R(N) obtained with the selection based on amplitude.}
\label{fig_glob2}
\end{figure}

\begin{figure}
\centering
\includegraphics[width=9cm]{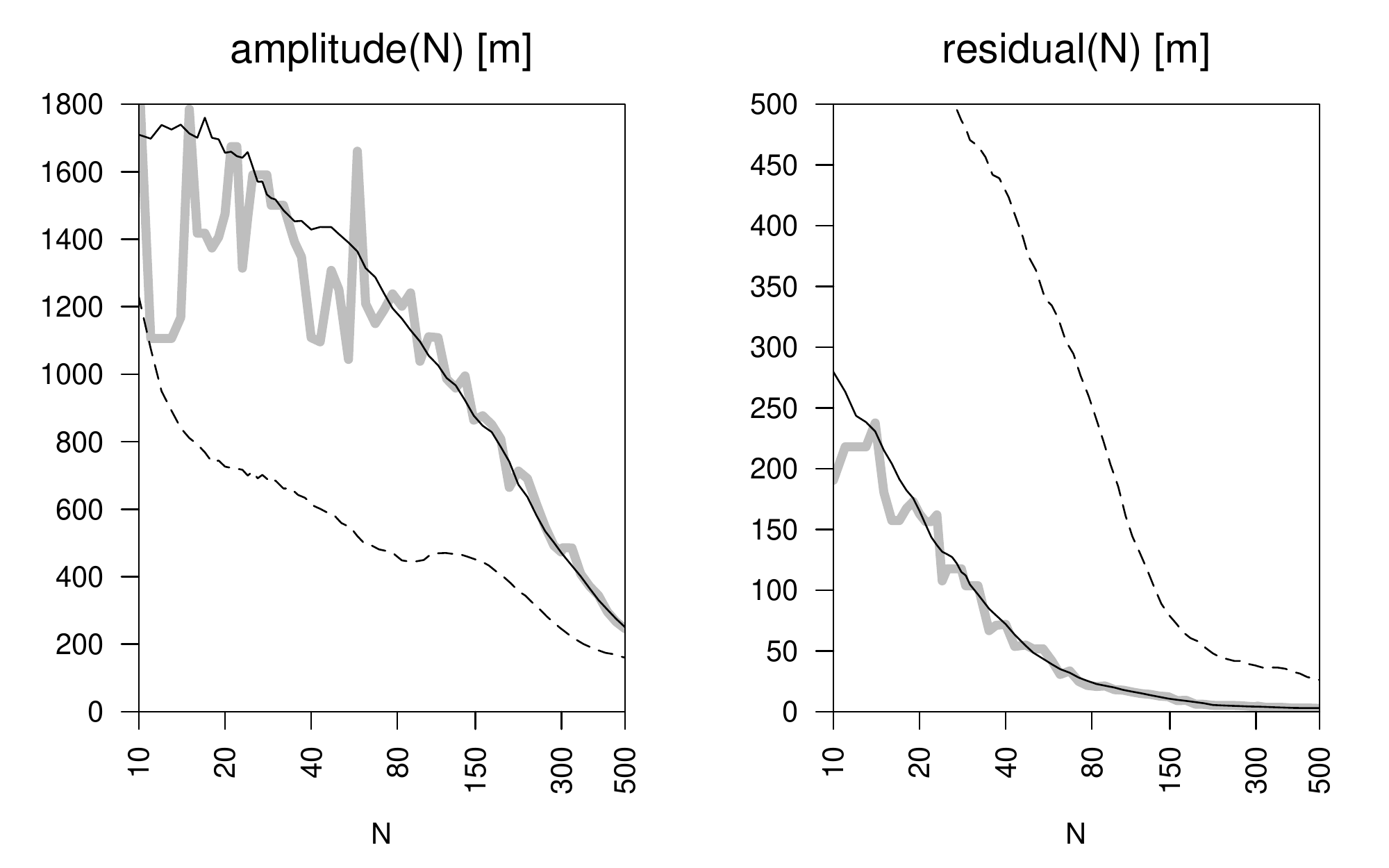}
\caption{Evolution with the MIQP selection of the amplitude of $\Delta D_{\mbox{\tiny{glob$^*$}}}$ and the amplitude of the residuals after fitting a ring for the standard set of masses (in gray) and corresponding averages computed over 100 random mass sets. The dashed lines represent the average amplitudes and residuals obtained with selection based on amplitude.}
\label{fig_ampli_res2}
\end{figure}

\begin{figure}
\centering
\includegraphics[width=9cm]{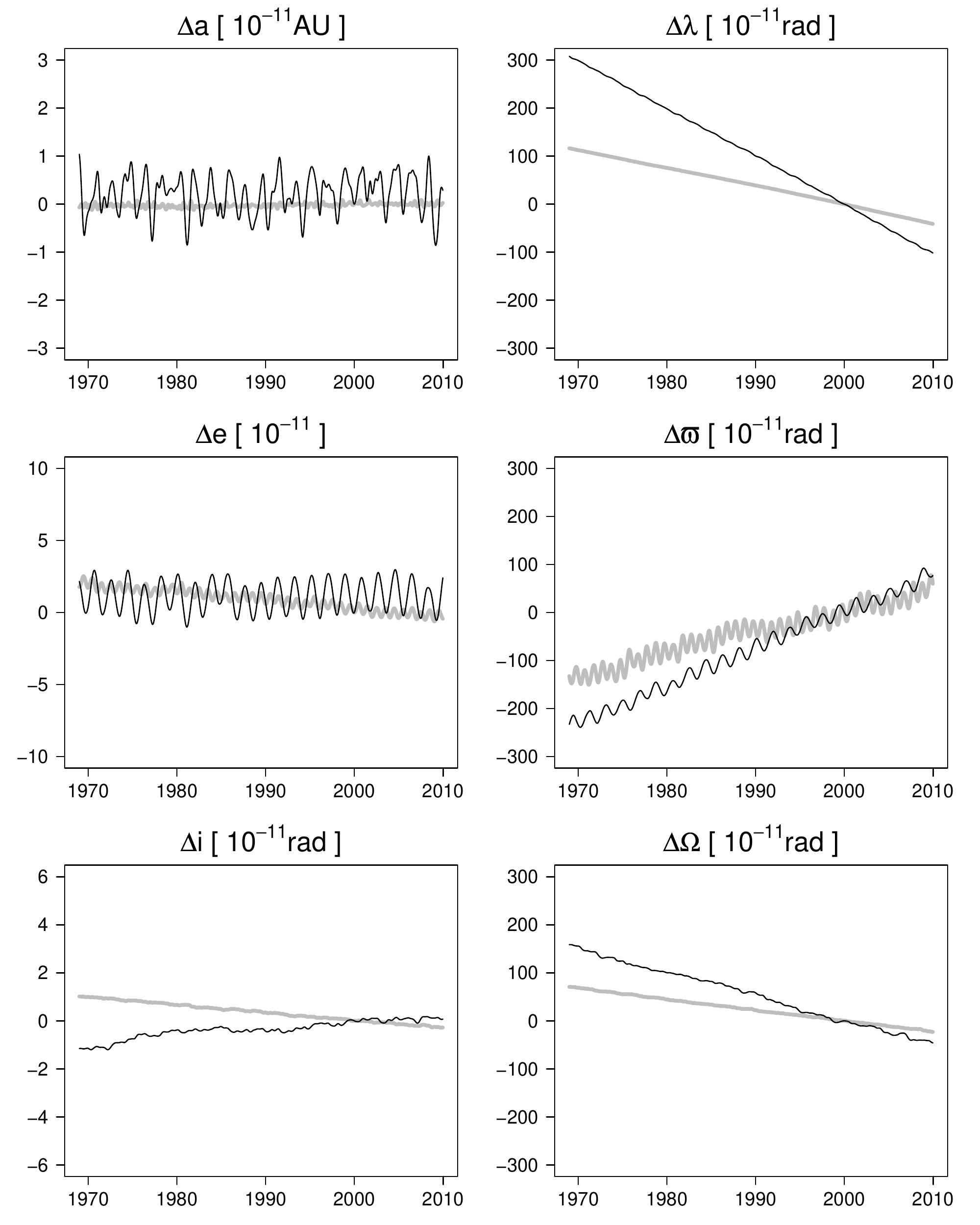}
\caption{Perturbation induced on the orbital elements of Earth (in gray) and Mars by the test model after removing from the test model at most 300 asteroids with the MIQP selection.}
\label{fig_Dx_glob2}
\end{figure}

The progressive removal of asteroids from the test model can be understood as a selection of individual asteroids that should be modeled individually in an ephemeris. The selection scheme based on the amplitude of the individual asteroid perturbations on the Earth-Mars distance is not optimal. Indeed for each state N, it is possible to slightly modify the set of the removed asteroids in order to eliminate the linear drift in the semi-major axis of Mars responsible for the quadratic evolution of the mean longitude in figure \ref{fig_Dx_glob}. The modification consists of removing from the test model a few additional asteroids that induce a positive slope in the perturbation of the semi-major axis of Mars and adding the same number of already removed objects with a negative slope. Such changes improve the average residuals from the previous 38 m to 20 m. 

A more systematic approach is to use a combinatorial optimization algorithm to select among the N most perturbing asteroids those that should be removed from the test model in order to maximize the modeling capacity of the ring. The problem can be stated formally for a particular mass set and a given N as the search for N+1 parameters $\alpha_i$ that minimize
\begin{eqnarray*}
\left\vert  \alpha_1 \Delta D_1 + ... + \alpha_N \Delta D_N + \alpha_{N+1} \Delta D_{\mbox{\tiny{ring}}} + \Delta D_{\mbox{\tiny{glob}}}(N)  \right\vert
\end{eqnarray*}
with the constraint that $\alpha_{N+1}$ is real positive and all the other $\alpha_i$ are binary (equal to 1 or 0). The problem falls into the category of mixed integer quadratic problems (MIQP). In this particular case, a direct search for a solution involves testing $2^N$ combinations and becomes very difficult for any useful value of N. The exact solution can nevertheless be found with various methods, \cite{gueye2009} provide a short literature review on the subject. One method of solving the MIQP is to linearize it and solve the linear formulation with the simplex algorithm provided for example in the GNU Linear Programming Kit (GLPK). A classical linearization is Glover's linearization \citep[see][]{gueye2009}, which involves N additional variables and 2N additional linear constraints. Unfortunately its implementation and the subsequent solution with GLPK does not lead to a solution within any reasonable time. Experiments with the commercial optimization package CPLEX\footnote[1]{The software solves a hierarchy of linear subproblems in a branch-and-bound approach, see www.ilog.com/products/cplex for more details on the package}, on the other hand, showed an ability to solve the quadratic problem within seconds for N below 120 and within minutes for N below 200. For higher values, the time taken to solve the problem becomes too large to be of practical use in our MC experiments. 

Figure \ref{fig_glob2} gives the analog of $R(N)$ obtained with a new selection scheme using the MIQP formulation, we denote the global effect corresponding to this new scheme with $\Delta D_{\mbox{\tiny{glob$^*$}}}(N)$. Because the calculations are relatively time consuming, only values of N below 500 are considered. For each N the algorithm uses CPLEX to select among the N most perturbing asteroids, those that should be removed from the test model to obtain an optimal fit with a ring. For N greater than 200, only the 200 least perturbing asteroids among the N most perturbing ones are considered by CPLEX. The remaining (N-200) are removed from the test model automatically. In figure \ref{fig_glob2} we observe that the ring's modeling capacity is greatly improved when compared with the selection scheme based on amplitude. Not only does $R(N)$ drop almost to zero, but the maximum modeling capacity is also reached earlier. Figure \ref{fig_ampli_res2} shows the evolution of the average maximum reached by the global perturbation and of the corresponding residuals after fitting the ring. Although the amplitude of the global effect is almost twice as large as in figure \ref{fig_ampli_res1}, the residuals are greatly improved. The performance reached after removing  the 300 most perturbing asteroids from the test model is obtained with the MIQP approach for $N = 50$. The average residuals after fitting the ring are 4 m for $N = 300$. This corresponds to approximately 1.3 m over 10 years. The unmodeled part of the global perturbation is thus below the one sigma residuals of the MGS/MO and MEX data obtained in INPOP08. The average maximum reached by the global effect is 472 m, which corresponds in terms of the ring's mass to $M_{\mbox{\tiny{ring}}} = 1.1 \pm 0.2\times 10^{-10} M_{\odot}$. 
 
We show in figure \ref{fig_Dx_glob2}, the perturbation induced on the orbital elements of Earth and Mars by the global effect obtained with the MIQP selection ($N = 300$) and the standard set of asteroid masses. The quadratic evolution of the mean longitude of Mars from figure \ref{fig_Dx_glob} is straightened up. This explains why the corresponding global effect on the Earth-Mars distance has increased in amplitude. The MIQP approach in general improves the resemblance between the effect on the orbital elements induced by the global perturbation and the ring. A comparison between figures \ref{fig_Dx_ring}, \ref{fig_Dx_glob}, and \ref{fig_Dx_glob2} shows that beside improvements in the semi-major axis and mean longitude perturbations, the new selection also improves the eccentricity and perihelion perturbations. There are still some discrepancies, most importantly in the perturbations induced on the inclinations and nodes as well as on the perihelium of the Earth; however, these discrepancies are not surprising as the asteroid selection is based only on the Earth-Mars distance, which according to Table \ref{tab_diff} is not very sensitive to these parameters.

\subsection{Accounting for all the inner planets} \label{sec_all}

\begin{figure}
\centering
\includegraphics[width=9cm]{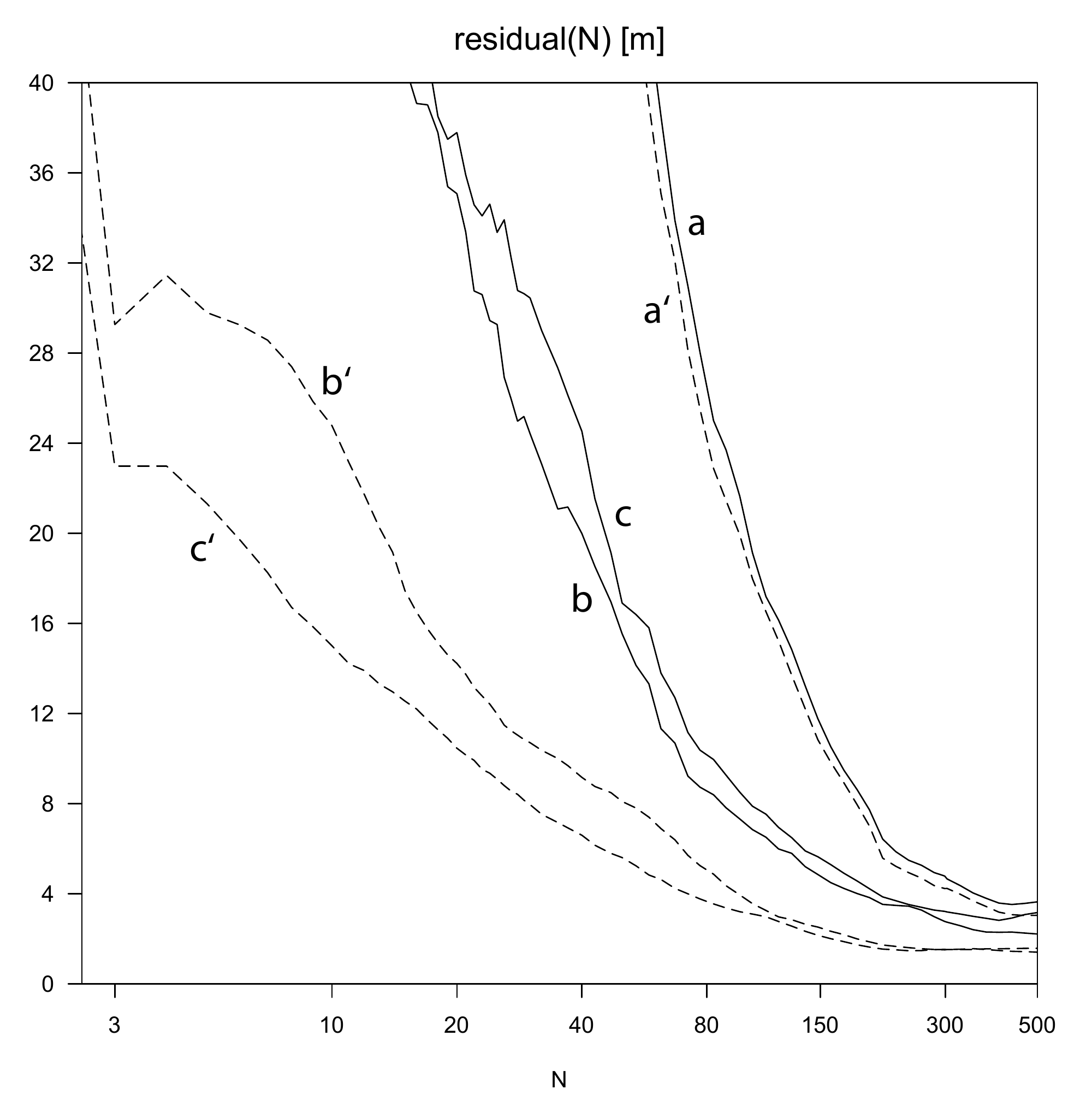}
\caption{Average residuals computed with the MIQP selection over 100 random mass sets for the Earth-Mars (a and a'), Earth-Venus (b and b'), and Earth-Mercury distances (c and c'). The continuous lines represent residuals obtained from a simultaneous fit of the ring to the $\Delta D_{\mbox{\tiny{glob$^*$}}}$ of the three planets, and the dashed lines represent the residuals obtained by fitting only one $\Delta D_{\mbox{\tiny{glob$^*$}}}$ at a time as in section \ref{sec_cplex}.}
\label{fig_res3_zoom}
\end{figure}

Today ephemerides are being fitted to accurate Venus ranging observations from the ongoing VEX mission \citep[see][]{fienga2009}. According to \cite{ashby2007}, Mercury ranging could become available within a few years with missions like Messenger or BepiColombo. To evaluate the ability of the ring to model the global effect on all the inner planets, we repeated the MC experiment made with the MIQP selection scheme in section \ref{sec_cplex}. Instead of fitting only the Earth-Mars distance, we simultaneously fit the effects on distances to Mercury, Venus, and Mars. Figure \ref{fig_res3_zoom} shows the evolution of residuals for all the planets in the simultaneous fit. For comparison the figure shows the residuals that can be obtained by fitting the distance to each planet separately. For $N = 300$, the ring is able to represent the global effect simultaneously on all the inner planets with an average accuracy better than 1.6 m over a 10 years time interval. Fitting all the planets together for $N = 300$ leads to the same amplitude of the global perturbation as in section \ref{sec_cplex}, and hence to the same estimate of the ring's mass. When considering separately Mercury or Venus, figure \ref{fig_res3_zoom} shows that residuals well below 1 m per year can be reached by removing merely 1 Ceres, 2 Pallas and 4 Vesta  from the main belt. 
 
\begin{table*}[t]
\begin{minipage}[t]{\textwidth}
\newcommand\T{\rule{0pt}{2.5ex}}
\newcommand\B{\rule[-1.5ex]{0pt}{0pt}}
\caption{Asteroids with an effect on the Earth-Mars distance greater than 100 m. The last column corresponds to the probability of being removed from the test model during the Monte Carlo experiments (see section \ref{ssec_ms}).} \label{tab_repro}
\centering\renewcommand{\arraystretch}{1.15}\renewcommand{\footnoterule}{}
\begin{tabular*}{18.5cm}{@{\extracolsep{\fill}} l r c r r r r}
\hline
\hline
Asteroid name \T \B & Diameter [km] & Density class\footnote{Density classes are assigned according to albedo and do not always correspond to taxonomies.} & Earth-Mars [m] & Earth-Venus [m] & Earth-Mercury [m] & prob. [$\%$]\\
\hline
4 Vesta & 468.30 & S & 11455.34 &  164.42 & 323.36 & 100  \\
1 Ceres & 848.40 & S &  7829.72 & 2656.87 & 935.08 & 100  \\
2 Pallas & 498.06 & S &  6612.41 &  977.16 & 379.08 & 100  \\
324 Bamberga & 229.44 & C &  1820.88 &   17.04 &  19.46 & 100  \\
10 Hygiea & 407.12 & C &   806.70 &  132.18 &  61.95 & 100  \\
6 Hebe & 185.18 & S &   691.70 &   54.66 &  21.15 & 100  \\
532 Herculina & 222.38 & S &   660.82 &   24.04 &   6.53 & 100  \\
19 Fortuna & 213.47 & C &   572.41 &   18.53 &   9.37 & 100  \\
51 Nemausa & 147.86 & M &   504.89 &   33.92 &  12.06 & 100  \\
7 Iris & 199.84 & S &   489.73 &   37.20 &  12.85 & 100  \\
9 Metis & 152.70 & S &   364.56 &    4.82 &   1.87 & 100  \\
20 Massalia & 145.50 & S &   285.84 &   20.90 &   7.31 & 100  \\
139 Juewa & 156.60 & C &   281.71 &    5.07 &   2.78 & 100  \\
31 Euphrosyne & 255.90 & C &   279.07 &   18.07 &   7.00 & 100  \\
16 Psyche & 253.16 & S &   264.08 &   38.66 &  13.92 & 100  \\
52 Europa & 302.50 & C &   252.60 &   43.76 &  15.93 & 100  \\
24 Themis & 218.44 & C &   234.96 &    2.07 &   1.25 & 100  \\
63 Ausonia & 103.14 & S &   234.40 &    4.15 &   1.44 & 100  \\
747 Winchester & 171.72 & C &   222.34 &   27.55 &  13.06 & 100  \\
60 Echo &  60.20 & S &   165.80 &    0.15 &   0.15 & 100  \\
78 Diana & 120.60 & C &   157.41 &    1.49 &   0.41 & 100  \\
354 Eleonora & 155.16 & S &   152.73 &    0.37 &   1.88 & 100  \\
41 Daphne & 174.00 & C &   149.44 &    6.11 &   4.25 & 100  \\
3 Juno & 233.92 & S &   145.29 &   15.88 &   5.58 & 100  \\
654 Zelinda & 127.40 & C &   142.71 &   21.79 &  13.53 & 100  \\
5 Astraea & 119.06 & S &   139.25 &   10.29 &   3.93 & 100  \\
18 Melpomene & 140.56 & S &   130.84 &   18.96 &  14.11 & 100  \\
128 Nemesis & 188.16 & C &   128.69 &   17.32 &   7.71 & 100  \\
192 Nausikaa & 103.26 & S &   128.46 &    0.67 &   2.26 & 100  \\
11 Parthenope & 153.34 & S &   115.67 &    5.73 &   3.86 & 100  \\
106 Dione & 146.60 & M &   114.27 &    4.99 &   1.58 & 100  \\
105 Artemis & 119.08 & C &   113.20 &    1.33 &   0.59 & 100  \\
23 Thalia & 107.54 & S &   112.92 &    1.05 &   0.56 & 100  \\
372 Palma & 188.62 & C &   112.84 &    3.79 &   1.58 & 100  \\
29 Amphitrite & 212.22 & S &   106.71 &   43.13 &  13.78 & 100  \\
419 Aurelia & 129.00 & C &   106.21 &    3.20 &   1.23 & 100  \\
15 Eunomia & 255.34 & S &   105.99 &   32.23 &  14.74 & 100  \\
14 Irene & 151.30 & S &   105.39 &   17.53 &   5.98 & 100  \\
48 Doris & 221.80 & C &   105.34 &    1.58 &   0.65 & 98  \\
488 Kreusa & 150.12 & C &   101.99 &    2.48 &   1.18 & 100  \\
\hline
\hline
\end{tabular*}
\end{minipage}
\end{table*}

\section{Applications} \label{sec_appli}

\subsection{Model selection}  \label{ssec_ms}

To model the global effect correctly with respect to the accuracy of available data, a reasonable choice according to figure \ref{fig_res3_zoom} is to account for fewer than 300 individual objects. By examining asteroids that were removed from the test model during the MC experiments, we can compile a list of objects that should be modeled individually in an ephemeris. For $N = 300$ in the simultaneous fit of section \ref{sec_all}, a total of 523 asteroids were removed at least once from the test model during the 100 MC runs, and the average number of removed objects was approximately 240. In an ephemeris with an idealized asteroid model we should therefore fit these 523 asteroid masses with the option of putting more than a half of the fitted values to zero. Among the 523 asteroids, there are 72 individuals removed from the test model on each run and 60 asteroids removed only once during the 100 runs. The distinction between asteroids having a high chance of removal and a low one is not clear, it is however possible to define an arbitrary limit above which the probability of being removed from the test model is reasonably high. Fixing this limit at $25\%$ leads to a total of 287 objects, which are listed in Table A1 of the supplementary online material. The table provides for each asteroid the probability of being selected as well as its maximum effects on the Earth-Mars, Earth-Venus, and Earth-Mercury distances during the 1969 - 2010 time interval (for the standard mass set). It is interesting to note the existence of asteroids with relatively small effects on the Earth-Mars distance but with very high probabilities of being included in the individual part of the asteroid model: for example, 758 Mancunia induces a perturbation of approximately 10 m, but it is removed on each of the 100 runs. A part of Table A1 of the supplementary online material is reproduced in Table \ref{tab_repro}. 


\subsection{Systematic error estimation} 

It is possible to use the results obtained in figure \ref{fig_res3_zoom} to estimate systematic errors that will be induced by the residuals of the global perturbation during future missions like BepiColombo. These systematic errors can have a significant impact on the planned determinations of physical parameters from the ranging data. An extensive study of this problem is presented in \cite{ashby2007}. 

The BepiColombo mission is expected to generate ranging data to Mercury accurate down to 4.5 cm \citep[see][]{ashby2007}. Figure \ref{fig_res3_zoom} shows that 4.5 cm per year is relatively close to the best possible residuals reached for the Earth-Mercury distance. To take full advantage of this accuracy, the asteroid model used to process the ranging data will have to correctly account for approximately 200 individual asteroids. This is a relatively high number because obtaining accurate estimates of 200 asteroid masses may still be difficult in the near future. If we estimate at 50 the number of asteroids that we are actually able to model with the highest accuracy, we can use figure \ref{fig_res3_zoom} to obtain an estimate of the systematic error for the BepiColombo mission if it took place today. For $N = 50$, fitting only the Earth-Mercury distance leads to residuals of approximately 6 m equivalent to a systematic error of 15 cm over a one year period. 

\section{Discussion}


It was shown that with an appropriate selection scheme a ring is able to effectively reduce the amplitude of a perturbation induced by thousands of asteroids from an average 472 m to only 4 m. The ring thus represents more than 99$\%$ of the global perturbation, which clearly makes it a very suitable model. The value of 99$\%$ was obtained with 100 MC experiments, so it can be considered as quite robust. Also, this percentage does not depend on choices made for attributing asteroids with masses in the test model. The estimations of the amplitude of the residuals, the amplitude of the global perturbation and the mass of the ring are, on the other hand, proportional to the mass of the test model and in consequence strongly dependent on the choices made in the attribution of asteroid masses. If for example in section \ref{ssec_astsmasses} the interval used for the C density class were centered on 1 instead of 1.5, all the previous parameters would have been approximately one third smaller. We are reasonably confident in the realism of the asteroid masses in our model because the mass of the ring obtained after removing the 300 most perturbing asteroids in section \ref{sec_comp1} is relatively close to the value of $0.34 \pm 0.15\times 10^{-10} M_{\odot}$ obtained in INPOP06. 

The objective in this work was to show that the ring is a first-order model of a main-belt global effect and that it is able to represent large numbers of objects in practice. The difficulty of fitting its mass with other highly correlated parameters is an important problem not considered here. In particular, the initial conditions of the planets were maintained fixed throughout our study, whereas they are fitted in an ephemeris. Because the global effect acts mostly through linear drifts in mean-longitudes, a large part can be absorbed by changes of a few meters in the initial semi-major axes of the planets.  The mass of the ring can be correlated with other parameters as well, like the individual asteroid masses or for example solar oblateness as shown by \cite{fienga2009b}. These correlations can be considered as so important that the ring is eventually not implemented in the model (the case in DE421) or its mass is fixed to a certain value (the case for INPOP08). The major arguments for keeping the ring in the model are its 99$\%$ modeling capacity and that, without the ring, systematic errors can reach several hundreds of meters.

In section \ref{ssec_analy0} we fixed the radius of the ring to 2.8 AU and always considered mass as the only parameter. We have briefly investigated the possibility of fitting the radius and mass together. We find out that, when considering only the Earth-Mars distance, any change in mass can effectively be compensated for by a change in radius. In terms of the residuals on the Earth-Mars distance, moving the ring from 2.8 AU to 2.4 AU is equivalent to doubling the ring's mass. Similarly moving the ring to 3.4 AU is equivalent to dividing the mass of the ring by two. The same residuals can be obtained no matter the radius. Nevertheless in section \ref{ssec_analy0}, the mass of a ring with radius 2.8 AU indeed corresponds to the total mass of the represented asteroids (it is not the case for other radii).

Throughout this paper, we estimated the amplitudes of perturbations with the maximum reached on the 1969 - 2010 interval. This corresponds to the maximum norm $\parallel.\parallel_\infty$. Measuring perturbations in terms of the root mean square (equivalent to the norm $\parallel.\parallel_2$) would divide all our amplitude estimations by approximately three. This would lead to much more relaxed demands on the asteroid models. In particular an accuracy below 2 m over 10 years would be reached in figure \ref{fig_res3_zoom} for $N = 100$ instead of $N = 300$. Similarly, the final systematic error for BepiColombo with a correct $N=50$ asteroid model would drop from 15 cm per year to 5 cm per year. We showed that the number of asteroids that need to be accounted for individually is no more than 300, if in a more optimistic perspective all our amplitude estimations can be divided by a factor of three, the number of asteroids to account for individually could be as low as 100. 

This study was restricted to the main-belt perturbations. Other Solar System objects potentially have impacts on the ephemerides that should be estimated and possibly accounted for. We can mention trans-Neptunian objects already implemented in the EPM ephemeris or the Trojan asteroids whose effect could be significant and which are certainly not accounted for by a ring. Also we only considered effects on the inner planets because they provide the majority of accurate data today. The effect of asteroids on the outer planets can be non-negligible and should be considered in future studies; especially, effects on Jupiter can be significant because of the various resonances with the main belt. However, such studies for the outer planets will have an impact on data only when accurate observations of the outer planets are available over a sufficient time span.

\section{Conclusion}

A ring is an implementation of an averaged orbit, which is a very good model of the perturbation induced by the main-belt asteroids. After removing less than 300 objects from the main-belt, the ring is able to account for more than 99$\%$ of the remaining perturbation on all the inner planets. Since the amplitude of the global effect can reach several hundreds of meters in terms of the Earth-Mars distance, it is advisable to keep the ring in an ephemeris model of the Solar System. 

\begin{appendix}

\section{3D perturbing force of an asteroid ring}
\label{app_force}

\begin{figure}
\centering
\includegraphics[width=8cm]{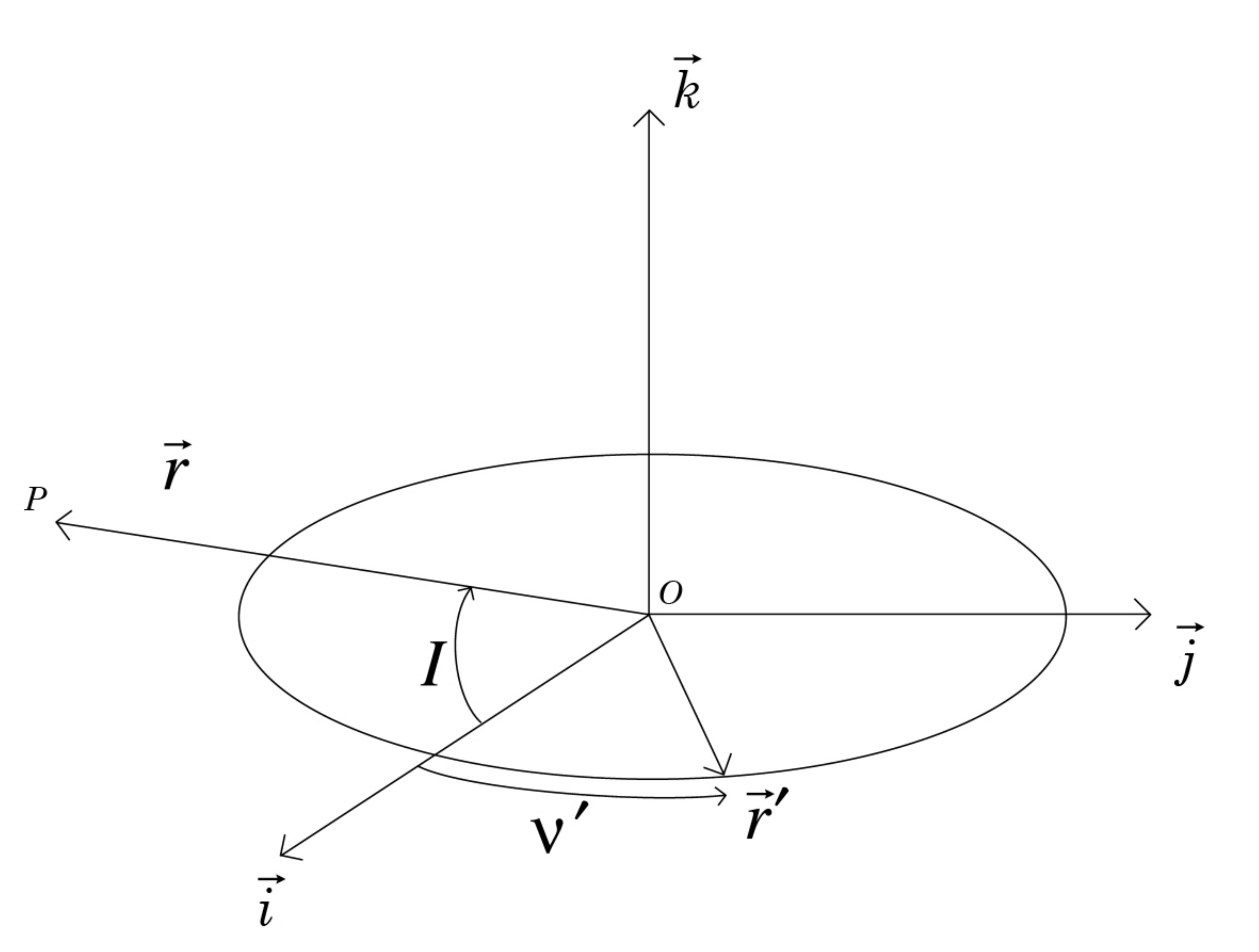}
\caption{The reference frame}
\label{fig_ref}
\end{figure}

\def\r{{\bf r}}
\def\i{{\bf i}}
\def\j{{\bf j}}
\def\k{{\bf k}}
\def\Frac#1#2{{{\displaystyle\strut#1}\over{\displaystyle\strut#2}}}
\def\norm#1{\left\Vert#1\right\Vert}
\newcommand{\g}{\gamma}
\newcommand{\cI}{{\cal I}}
\newcommand{\cK}{{\cal K}}
\newcommand{\cE}{{\cal E}}

We derive here the expression of the gravitational force exerted by a ring of center $O$, radius $r'$, and mass $m$ on a point $P$ of mass $M$. The chosen reference frame is $(O, \i, \j, \k)$, with $\k$ orthogonal to the ring's plane and $\i$ the unit vector in the direction of the projection of $\r=\overrightarrow{OP}$ on the ring's plane. We denote with $\r'$ the radius vector of a point on the ring. The longitude angle of $\r'$ with origin at $\i$ is denoted $v'$ (see figure \ref{fig_ref}). The potential exerted by the ring on $P$ is 
\begin{eqnarray*}
V(\r) = -\Frac{GmM}{2\pi}\int_0^{2\pi} \Frac{dv'}{\norm{\r'-\r}} .
\end{eqnarray*}
By defining $I$ as the angle of $\r$ with the plane of the ring, the potential can be rewritten as
\begin{eqnarray*}
V(\r) = -\Frac{GmM}{2\pi}\int_0^{2\pi} \Frac{dv'}{(r^2+r'^2-2rr'\cos I\cos v')^{1/2}} .
\end{eqnarray*}
With 
\begin{eqnarray*}
\alpha &=& \Frac{2rr'\cos I}{r^2+r'^2} ,
\end{eqnarray*}
we obtain
\begin{eqnarray*}
V(\r) &=&-\Frac{GmM}{\sqrt{r^2+r'^2}}\Frac{1}{2\pi}\int_0^{2\pi} \Frac{dv'}{(1-\alpha\cos v')^{1/2}} .
\end{eqnarray*}
The above expression can be rewritten as 
\begin{eqnarray*}
V(\r) = -\Frac{GmM}{\sqrt{\Gamma}}\Frac{2}{\pi}\cK(\beta) ,
\end{eqnarray*}
where $\cK(\beta)$ is the complete elliptical integral\footnote[1]{The elliptical integral of first ($\cK$) and second ($\cE$) 
kind are defined as \citep{whittaker1927} :
\begin{eqnarray*}
\cK(\beta) = \int_0^{\pi/2} \Frac{1}{\sqrt{1-\beta\sin^2 x}} dx \ ;\quad
\cE(\beta) = \int_0^{\pi/2} \sqrt{1-\beta\sin^2 x} dx.
\end{eqnarray*}
}
of the first kind, and $\beta$ and $\Gamma$ are defined as
\begin{eqnarray*}
\beta = \Frac{2\alpha}{1+\alpha} \ \quad
\Gamma = r^2+r'^2+2r'r\cos I .
\end{eqnarray*}
The force exerted on  $P$ is ${\bf F} = -\nabla_\r V$. If 
\begin{eqnarray*}
\nabla_\r(\Gamma) = 2\r + 2r'\i \ \quad 
\nabla_\r(\beta) =  -\Frac{8r'r\cos I }{\Gamma^2} \r + 4r'\Frac{r^2+r'^2}{\Gamma^2}\i  ,
\end{eqnarray*}
we obtain, after straightforward computation, the expression of ${\bf F}$ in terms of the complete elliptical integral of first ($\cK$) and second ($\cE$) kind :
\begin{eqnarray*}
{\bf F} = -\Frac{2 GmM}{\pi \alpha(1-\beta)\Gamma^{3/2}} \left[ \alpha\cE(\beta) \r
+\left( ( 1-\alpha)\cK(\beta) -\cE(\beta)\right) r'\i\right] .
\end{eqnarray*}
This expression is valid for an internal or external body $P$ with $r\neq r'$. Indeed in this case we always have $0 \leq \alpha < 1 $, hence also $0 \leq \beta < 1 $, so the elliptical integrals are well defined by their hypergeometric series, although one may use approximation formulas for a fast evaluation. When the problem is planar ($I=0$), these expressions become equivalent to the expressions given in \citet{krasinsky2002} or \citet{fienga2008}.

\end{appendix}

\begin{acknowledgements}
The first author wishes to acknowledge interesting discussions with James Hilton (US Naval Observatory). This work was done with the financial support of the CNES and the French Ministry of Education.
\end{acknowledgements}

\bibliographystyle{aa.bst}
\bibliography{article}

\renewcommand{\thetable}{A\arabic{table}}

\onllongtab{1}{
\newcommand\T{\rule{0pt}{2.5ex}}
\newcommand\B{\rule[-1.5ex]{0pt}{0pt}}
\renewcommand{\arraystretch}{1.15}
\begin{longtable}{l r c r r r r}
\caption{Asteroids selected for the individual part of the asteroid model when simultaneously fitting ranging data on all four inner planets. As described in section \ref{sec_appli}, only asteroids with probability greater than $25\%$ of being removed from the test model during the 100 MC experiments are listed. For each asteroid this probabilty is given together with the asteroid's diameter, density class (in the standard mass set) and correponding perturbations on the Earth-Mars, Earth-Venus, and Earth-Mercury distances in terms of max $\vert \Delta D \vert$ on the 1969-2010 time interval. Density classes are assigned according to albedo (see section \ref{ssec_astsmasses}) and do not always correspond to taxonomies.}\\
\hline
\hline
Asteroid name \T \B & Diameter [km] & Density class & Earth-Mars [m] & Earth-Venus [m] & Earth-Mercury [m] & prob. [$\%$]\\
\hline
\endfirsthead
\caption{Continued.} \\
\hline
Asteroid name \T \B& Diameter [km] & Density class & Earth-Mars [m] & Earth-Venus [m] & Earth-Mercury [m] & prob. [$\%$]\\
\hline
\endhead
\hline
\endfoot
\hline
\endlastfoot
4 Vesta & 468.30 & S & 11455.34 &  164.42 & 323.36 & 100  \\
1 Ceres & 848.40 & S &  7829.72 & 2656.87 & 935.08 & 100  \\
2 Pallas & 498.06 & S &  6612.41 &  977.16 & 379.08 & 100  \\
324 Bamberga & 229.44 & C &  1820.88 &   17.04 &  19.46 & 100  \\
10 Hygiea & 407.12 & C &   806.70 &  132.18 &  61.95 & 100  \\
6 Hebe & 185.18 & S &   691.70 &   54.66 &  21.15 & 100  \\
532 Herculina & 222.38 & S &   660.82 &   24.04 &   6.53 & 100  \\
19 Fortuna & 213.47 & C &   572.41 &   18.53 &   9.37 & 100  \\
51 Nemausa & 147.86 & M &   504.89 &   33.92 &  12.06 & 100  \\
7 Iris & 199.84 & S &   489.73 &   37.20 &  12.85 & 100  \\
9 Metis & 152.70 & S &   364.56 &    4.82 &   1.87 & 100  \\
20 Massalia & 145.50 & S &   285.84 &   20.90 &   7.31 & 100  \\
139 Juewa & 156.60 & C &   281.71 &    5.07 &   2.78 & 100  \\
31 Euphrosyne & 255.90 & C &   279.07 &   18.07 &   7.00 & 100  \\
16 Psyche & 253.16 & S &   264.08 &   38.66 &  13.92 & 100  \\
52 Europa & 302.50 & C &   252.60 &   43.76 &  15.93 & 100  \\
24 Themis & 218.44 & C &   234.96 &    2.07 &   1.25 & 100  \\
63 Ausonia & 103.14 & S &   234.40 &    4.15 &   1.44 & 100  \\
747 Winchester & 171.72 & C &   222.34 &   27.55 &  13.06 & 100  \\
60 Echo &  60.20 & S &   165.80 &    0.15 &   0.15 & 100  \\
78 Diana & 120.60 & C &   157.41 &    1.49 &   0.41 & 100  \\
354 Eleonora & 155.16 & S &   152.73 &    0.37 &   1.88 & 100  \\
41 Daphne & 174.00 & C &   149.44 &    6.11 &   4.25 & 100  \\
3 Juno & 233.92 & S &   145.29 &   15.88 &   5.58 & 100  \\
654 Zelinda & 127.40 & C &   142.71 &   21.79 &  13.53 & 100  \\
5 Astraea & 119.06 & S &   139.25 &   10.29 &   3.93 & 100  \\
18 Melpomene & 140.56 & S &   130.84 &   18.96 &  14.11 & 100  \\
128 Nemesis & 188.16 & C &   128.69 &   17.32 &   7.71 & 100  \\
192 Nausikaa & 103.26 & S &   128.46 &    0.67 &   2.26 & 100  \\
11 Parthenope & 153.34 & S &   115.67 &    5.73 &   3.86 & 100  \\
106 Dione & 146.60 & M &   114.27 &    4.99 &   1.58 & 100  \\
105 Artemis & 119.08 & C &   113.20 &    1.33 &   0.59 & 100  \\
23 Thalia & 107.54 & S &   112.92 &    1.05 &   0.56 & 100  \\
372 Palma & 188.62 & C &   112.84 &    3.79 &   1.58 & 100  \\
29 Amphitrite & 212.22 & S &   106.71 &   43.13 &  13.78 & 100  \\
419 Aurelia & 129.00 & C &   106.21 &    3.20 &   1.23 & 100  \\
15 Eunomia & 255.34 & S &   105.99 &   32.23 &  14.74 & 100  \\
14 Irene & 151.30 & S &   105.39 &   17.53 &   5.98 & 100  \\
48 Doris & 221.80 & C &   105.34 &    1.58 &   0.65 & 98  \\
488 Kreusa & 150.12 & C &   101.99 &    2.48 &   1.18 & 100  \\
451 Patientia & 224.96 & C &    99.02 &    4.18 &   1.03 & 99  \\
344 Desiderata & 132.28 & C &    96.77 &    2.26 &   1.94 & 100  \\
109 Felicitas &  89.44 & C &    95.92 &    1.79 &   1.08 & 98  \\
405 Thia & 124.90 & C &    95.73 &    5.82 &   3.21 & 100  \\
511 Davida & 326.06 & C &    87.22 &   12.60 &   2.87 & 100  \\
129 Antigone & 161.20 & M &    80.69 &   26.62 &   6.71 & 93  \\
83 Beatrix &  81.38 & M &    78.82 &    2.93 &   1.68 & 86  \\
88 Thisbe & 200.58 & C &    76.54 &   14.11 &   2.41 & 99  \\
145 Adeona & 151.14 & C &    71.58 &    2.53 &   2.96 & 87  \\
187 Lamberta & 130.40 & C &    70.84 &    9.56 &   2.01 & 94  \\
98 Ianthe & 104.46 & C &    70.43 &    2.10 &   0.83 & 99  \\
13 Egeria & 207.64 & C &    68.44 &    4.06 &   2.41 & 97  \\
27 Euterpe & 109.61 & S &    67.04 &    2.09 &   0.49 & 100  \\
25 Phocaea &  75.12 & S &    65.91 &    1.73 &   0.32 & 100  \\
240 Vanadis & 103.90 & C &    64.32 &    1.94 &   0.92 & 100  \\
356 Liguria & 131.32 & C &    63.12 &    1.22 &   0.49 & 100  \\
230 Athamantis & 108.98 & S &    63.11 &    1.44 &   0.35 & 100  \\
22 Kalliope & 181.00 & S &    61.26 &    6.96 &   1.05 & 100  \\
704 Interamnia & 316.62 & C &    61.16 &    7.87 &   6.50 & 100  \\
505 Cava & 107.98 & C &    61.10 &    1.65 &   0.73 & 100  \\
194 Prokne & 168.42 & C &    60.99 &    4.43 &   3.98 & 99  \\
56 Melete & 113.24 & C &    59.77 &    4.60 &   1.62 & 99  \\
268 Adorea & 139.88 & C &    58.53 &    4.58 &   1.53 & 79  \\
53 Kalypso & 115.38 & C &    56.04 &    0.50 &   0.36 & 99  \\
585 Bilkis &  58.10 & C &    55.47 &    0.14 &   0.20 & 75  \\
94 Aurora & 204.88 & C &    54.04 &    4.16 &   3.04 & 85  \\
74 Galatea & 118.70 & C &    53.67 &    1.08 &   0.78 & 87  \\
259 Aletheia & 178.60 & C &    53.47 &    1.90 &   1.76 & 100  \\
12 Victoria & 112.76 & S &    50.57 &   12.31 &   3.76 & 100  \\
42 Isis & 100.20 & S &    50.15 &    0.70 &   0.50 & 100  \\
141 Lumen & 131.04 & C &    49.65 &   10.11 &   5.24 & 98  \\
8 Flora & 135.88 & S &    48.72 &    3.24 &   3.13 & 100  \\
393 Lampetia &  96.90 & C &    46.80 &    1.89 &   1.54 & 99  \\
81 Terpsichore & 119.08 & C &    46.67 &    5.81 &   3.29 & 94  \\
87 Sylvia & 260.94 & C &    46.17 &   15.28 &   4.24 & 100  \\
40 Harmonia & 107.62 & S &    45.73 &    0.57 &   1.49 & 100  \\
694 Ekard &  90.78 & C &    43.86 &    0.84 &   0.74 & 100  \\
37 Fides & 108.34 & S &    43.28 &    0.31 &   0.72 & 100  \\
144 Vibilia & 142.38 & C &    43.26 &    1.75 &   1.05 & 99  \\
36 Atalante & 105.60 & C &    42.90 &    1.84 &   0.83 & 84  \\
410 Chloris & 123.56 & C &    41.81 &    0.85 &   0.24 & 90  \\
313 Chaldaea &  96.34 & C &    39.34 &    0.55 &   0.92 & 93  \\
50 Virginia &  99.82 & C &    39.16 &    6.28 &   2.36 & 98  \\
173 Ino & 154.10 & C &    39.14 &    2.81 &   0.54 & 100  \\
449 Hamburga &  85.60 & C &    39.06 &    0.85 &   0.15 & 95  \\
516 Amherstia &  73.10 & S &    38.66 &    0.53 &   0.24 & 99  \\
387 Aquitania & 100.52 & S &    38.34 &    4.69 &   2.14 & 92  \\
164 Eva & 104.88 & C &    36.51 &    1.57 &   0.94 & 98  \\
59 Elpis & 164.80 & C &    36.49 &    8.28 &   1.84 & 72  \\
107 Camilla & 222.62 & C &    36.35 &   13.62 &   4.20 & 93  \\
30 Urania & 100.16 & S &    35.99 &    1.26 &   0.14 & 100  \\
247 Eukrate & 134.44 & C &    35.46 &    0.57 &   0.73 & 66  \\
409 Aspasia & 161.62 & C &    35.33 &    6.27 &   2.04 & 93  \\
17 Thetis &  90.04 & S &    35.24 &    4.19 &   1.59 & 81  \\
386 Siegena & 165.00 & C &    34.72 &    2.05 &   0.75 & 100  \\
49 Pales & 149.80 & C &    34.10 &    9.27 &   3.36 & 97  \\
423 Diotima & 208.78 & C &    33.98 &   10.06 &   3.21 & 100  \\
97 Klotho &  82.82 & S &    33.84 &    0.35 &   0.44 & 94  \\
404 Arsinoe &  97.70 & C &    33.43 &    0.99 &   0.39 & 66  \\
1021 Flammario &  99.40 & C &    32.21 &    4.11 &   3.02 & 76  \\
46 Hestia & 124.14 & C &    31.98 &    2.67 &   1.65 & 60  \\
135 Hertha &  79.24 & S &    31.96 &    0.67 &   1.10 & 80  \\
115 Thyra &  79.82 & S &    31.89 &    0.66 &   0.13 & 100  \\
521 Brixia & 115.66 & C &    31.60 &    6.43 &   3.65 & 100  \\
444 Gyptis & 163.08 & C &    31.31 &    4.86 &   2.37 & 100  \\
212 Medea & 136.12 & C &    30.92 &    4.67 &   1.36 & 85  \\
85 Io & 154.78 & C &    30.56 &    5.84 &   4.05 & 99  \\
304 Olga &  67.86 & C &    30.50 &    0.17 &   0.14 & 38  \\
21 Lutetia &  95.76 & S &    30.47 &    2.31 &   0.68 & 98  \\
181 Eucharis & 106.66 & S &    30.32 &    2.54 &   0.86 & 99  \\
89 Julia & 151.46 & S &    30.23 &   17.04 &   9.29 & 77  \\
111 Ate & 134.56 & C &    29.28 &    0.53 &   0.18 & 91  \\
471 Papagena & 134.18 & S &    28.89 &    4.00 &   0.90 & 85  \\
96 Aegle & 170.02 & C &    28.86 &    3.46 &   2.34 & 99  \\
44 Nysa &  70.64 & S &    28.79 &    2.58 &   0.94 & 99  \\
416 Vaticana &  85.46 & S &    28.74 &    0.96 &   0.31 & 79  \\
335 Roberta &  89.06 & C &    28.58 &    0.96 &   0.31 & 83  \\
150 Nuwa & 151.12 & C &    28.09 &    2.91 &   0.26 & 84  \\
275 Sapientia &  96.68 & S &    27.62 &    0.87 &   0.11 & 52  \\
216 Kleopatra & 135.06 & S &    27.33 &   13.77 &   4.38 & 100  \\
69 Hesperia & 138.12 & S &    26.99 &    5.74 &   2.24 & 97  \\
469 Argentina & 125.56 & C &    26.36 &    1.38 &   0.66 & 48  \\
365 Corduba & 105.92 & C &    26.06 &    3.75 &   1.35 & 92  \\
593 Titania &  75.32 & C &    26.01 &    0.10 &   0.12 & 76  \\
680 Genoveva &  83.92 & C &    25.67 &    0.86 &   0.31 & 99  \\
554 Peraga &  95.88 & C &    25.41 &    7.98 &   3.31 & 91  \\
75 Eurydike &  55.90 & S &    25.22 &    0.23 &   0.13 & 97  \\
130 Elektra & 182.24 & C &    25.12 &    4.74 &   2.04 & 97  \\
38 Leda & 115.94 & C &    23.89 &    2.44 &   1.07 & 99  \\
200 Dynamene & 128.36 & C &    23.33 &    3.63 &   1.58 & 100  \\
481 Emita & 108.47 & C &    23.02 &    0.69 &   0.22 & 98  \\
914 Palisana &  76.60 & M &    22.97 &    3.39 &   0.89 & 94  \\
760 Massinga &  71.30 & S &    22.48 &    0.93 &   0.29 & 94  \\
163 Erigone &  72.62 & C &    22.43 &    1.31 &   0.64 & 95  \\
385 Ilmatar &  91.54 & S &    22.36 &    3.52 &   1.13 & 51  \\
93 Minerva & 141.56 & C &    22.28 &    1.96 &   0.45 & 53  \\
788 Hohensteina & 103.68 & C &    21.88 &    0.56 &   0.38 & 100  \\
776 Berbericia & 151.16 & C &    21.86 &    2.46 &   0.53 & 62  \\
702 Alauda & 194.72 & C &    21.38 &    8.03 &   3.15 & 95  \\
193 Ambrosia &  65.97 & S &    21.31 &    1.31 &   1.00 & 37  \\
455 Bruchsalia &  84.40 & C &    20.54 &    1.27 &   0.54 & 97  \\
211 Isolda & 143.18 & C &    20.41 &    1.41 &   0.17 & 90  \\
238 Hypatia & 148.50 & C &    19.91 &    6.49 &   1.94 & 84  \\
762 Pulcova & 137.08 & C &    19.90 &    0.59 &   0.28 & 100  \\
132 Aethra &  42.86 & S &    19.75 &    0.48 &   0.29 & 83  \\
814 Tauris & 109.56 & C &    19.71 &    0.78 &   1.10 & 82  \\
415 Palatia &  76.34 & C &    19.51 &    0.62 &   0.57 & 97  \\
626 Notburga & 100.74 & C &    19.20 &    4.48 &   1.87 & 78  \\
674 Rachele &  97.34 & S &    18.85 &    2.96 &   1.12 & 64  \\
28 Bellona & 120.90 & S &    18.81 &    1.27 &   0.68 & 100  \\
68 Leto & 122.56 & S &    18.51 &    3.28 &   1.59 & 99  \\
70 Panopaea & 122.18 & C &    18.38 &    1.61 &   0.76 & 97  \\
751 Faina & 110.50 & C &    18.38 &    1.76 &   0.39 & 68  \\
752 Sulamitis &  62.78 & C &    18.38 &    0.48 &   0.21 & 26  \\
287 Nephthys &  67.60 & S &    18.22 &    1.73 &   0.61 & 95  \\
91 Aegina & 109.82 & C &    17.97 &    3.53 &   1.70 & 63  \\
346 Hermentaria & 106.52 & S &    17.76 &    4.07 &   1.98 & 84  \\
895 Helio & 141.90 & C &    17.48 &    1.22 &   0.34 & 97  \\
322 Phaeo &  70.84 & C &    16.98 &    0.28 &   0.14 & 46  \\
779 Nina &  76.62 & S &    16.92 &    0.26 &   0.45 & 89  \\
350 Ornamenta & 118.34 & C &    16.55 &    0.47 &   0.18 & 98  \\
34 Circe & 113.54 & C &    15.70 &    0.14 &   0.11 & 49  \\
602 Marianna & 124.72 & C &    15.56 &    0.24 &   0.40 & 94  \\
712 Boliviana & 127.56 & C &    14.91 &    0.84 &   1.41 & 96  \\
325 Heidelberga &  75.72 & M &    14.75 &    1.54 &   0.50 & 87  \\
786 Bredichina &  91.60 & C &    14.65 &    0.69 &   0.11 & 34  \\
80 Sappho &  78.40 & S &    14.59 &    3.96 &   1.58 & 76  \\
84 Klio &  79.16 & C &    14.36 &    0.80 &   0.29 & 86  \\
308 Polyxo & 140.68 & C &    14.24 &    0.47 &   0.22 & 60  \\
209 Dido & 159.94 & C &    14.02 &    4.95 &   2.13 & 59  \\
241 Germania & 168.90 & C &    14.02 &    2.63 &   2.01 & 87  \\
236 Honoria &  86.20 & S &    13.91 &    0.89 &   0.26 & 73  \\
804 Hispania & 157.58 & C &    13.89 &    5.51 &   1.67 & 75  \\
912 Maritima &  83.18 & M &    13.81 &    1.96 &   1.05 & 82  \\
337 Devosa &  59.10 & S &    13.75 &    1.27 &   0.38 & 96  \\
54 Alexandra & 165.76 & C &    13.63 &    9.30 &   3.57 & 93  \\
336 Lacadiera &  69.32 & C &    13.58 &    0.84 &   0.21 & 90  \\
784 Pickeringia &  89.42 & C &    13.49 &    0.30 &   0.04 & 93  \\
121 Hermione & 209.00 & C &    13.48 &    3.85 &   2.49 & 67  \\
118 Peitho &  41.72 & S &    13.25 &    0.19 &   0.09 & 82  \\
266 Aline & 109.10 & C &    12.98 &    0.08 &   0.42 & 54  \\
517 Edith &  91.12 & C &    12.24 &    0.17 &   0.07 & 89  \\
72 Feronia &  85.90 & C &    12.21 &    1.79 &   0.56 & 38  \\
595 Polyxena & 109.06 & M &    12.00 &    2.94 &   1.01 & 37  \\
120 Lachesis & 174.10 & C &    11.81 &    4.60 &   0.30 & 71  \\
599 Luisa &  64.88 & S &    11.80 &    1.97 &   0.95 & 74  \\
112 Iphigenia &  72.18 & C &    11.71 &    1.31 &   0.73 & 48  \\
61 Danae &  82.04 & S &    11.51 &    1.20 &   0.41 & 80  \\
442 Eichsfeldia &  66.74 & C &    11.42 &    1.49 &   0.48 & 50  \\
485 Genua &  63.88 & S &    11.30 &    0.29 &   0.14 & 87  \\
45 Eugenia & 214.62 & C &    11.04 &   11.47 &   4.07 & 100  \\
582 Olympia &  43.40 & S &    11.04 &    0.17 &   0.08 & 48  \\
39 Laetitia & 149.52 & S &    10.75 &    1.65 &   2.26 & 98  \\
568 Cheruskia &  86.98 & C &    10.74 &    2.56 &   0.99 & 62  \\
156 Xanthippe & 120.98 & C &    10.70 &    5.90 &   1.03 & 83  \\
148 Gallia &  97.76 & S &    10.53 &    0.27 &   0.17 & 93  \\
1171 Rusthawelia &  70.12 & C &    10.43 &    0.39 &   0.17 & 47  \\
134 Sophrosyne & 123.26 & C &    10.30 &    1.70 &   0.27 & 84  \\
758 Mancunia &  85.48 & S &    10.30 &    0.48 &   0.17 & 100  \\
345 Tercidina &  94.12 & C &    10.25 &    0.58 &   0.12 & 73  \\
667 Denise &  81.28 & C &    10.13 &    0.09 &   0.05 & 57  \\
323 Brucia &  35.82 & S &    10.10 &    0.15 &   0.07 & 63  \\
92 Undina & 126.42 & S &     9.91 &    3.70 &   1.96 & 44  \\
79 Eurynome &  66.46 & S &     9.86 &    0.44 &   0.17 & 41  \\
221 Eos & 103.88 & S &     9.79 &    1.60 &   0.51 & 93  \\
283 Emma & 148.06 & C &     9.79 &    5.34 &   1.81 & 81  \\
503 Evelyn &  81.68 & C &     9.67 &    0.09 &   0.05 & 50  \\
618 Elfriede & 120.30 & C &     9.63 &    0.98 &   0.42 & 29  \\
47 Aglaja & 126.96 & C &     9.54 &    1.96 &   1.46 & 66  \\
172 Baucis &  62.42 & S &     9.43 &    0.09 &   0.14 & 95  \\
77 Frigga &  69.24 & S &     9.37 &    0.75 &   0.36 & 97  \\
566 Stereoskopia & 168.16 & C &     9.28 &    1.37 &   0.47 & 80  \\
217 Eudora &  66.24 & C &     9.24 &    0.50 &   0.27 & 77  \\
171 Ophelia & 116.68 & C &     9.06 &    3.65 &   1.21 & 71  \\
739 Mandeville & 107.54 & C &     8.94 &    0.39 &   0.46 & 69  \\
790 Pretoria & 170.38 & C &     8.83 &    0.43 &   0.37 & 51  \\
147 Protogeneia & 132.94 & C &     8.64 &    3.24 &   0.95 & 58  \\
86 Semele & 120.56 & C &     8.25 &    1.28 &   0.78 & 30  \\
62 Erato &  95.40 & C &     8.24 &    0.75 &   0.23 & 70  \\
250 Bettina &  79.76 & S &     8.24 &    1.12 &   0.41 & 56  \\
71 Niobe &  83.42 & S &     8.09 &    0.26 &   0.25 & 49  \\
185 Eunike & 157.50 & C &     7.88 &    5.32 &   2.08 & 43  \\
591 Irmgard &  51.86 & C &     7.80 &    0.10 &   0.06 & 36  \\
234 Barbara &  43.76 & S &     7.69 &    0.16 &   0.06 & 93  \\
978 Aidamina &  78.74 & C &     7.68 &    0.16 &   0.06 & 47  \\
679 Pax &  51.46 & S &     7.64 &    1.02 &   0.34 & 83  \\
420 Bertholda & 141.24 & C &     7.39 &    3.05 &   1.02 & 57  \\
95 Arethusa & 136.04 & C &     7.35 &    0.72 &   0.50 & 63  \\
735 Marghanna &  74.32 & C &     7.34 &    2.80 &   1.58 & 63  \\
769 Tatjana & 106.44 & C &     7.23 &    0.30 &   0.14 & 57  \\
210 Isabella &  86.66 & C &     7.15 &    0.48 &   0.11 & 42  \\
117 Lomia & 148.72 & C &     7.09 &    2.50 &   0.30 & 45  \\
491 Carina &  97.30 & C &     7.08 &    1.23 &   0.54 & 64  \\
1963 Bezovec &  44.68 & C &     6.96 &    0.09 &   0.02 & 62  \\
122 Gerda &  81.68 & S &     6.88 &    0.59 &   0.23 & 77  \\
196 Philomela & 136.38 & S &     6.88 &    5.20 &   1.88 & 54  \\
388 Charybdis & 114.18 & C &     6.84 &    1.39 &   0.11 & 52  \\
223 Rosa &  87.60 & C &     6.73 &    0.89 &   0.50 & 58  \\
1107 Lictoria &  79.18 & C &     6.73 &    0.12 &   0.06 & 41  \\
431 Nephele &  95.04 & C &     6.69 &    0.87 &   0.38 & 68  \\
168 Sibylla & 148.38 & C &     6.65 &    1.04 &   0.27 & 41  \\
1036 Ganymed &  31.66 & S &     6.49 &    6.92 &   3.74 & 76  \\
138 Tolosa &  45.50 & S &     6.43 &    0.37 &   0.14 & 73  \\
949 Hel &  69.18 & C &     6.34 &    0.33 &   0.11 & 35  \\
893 Leopoldina &  76.14 & C &     6.29 &    0.70 &   0.26 & 49  \\
179 Klytaemnestra &  77.68 & S &     6.27 &    0.88 &   0.34 & 78  \\
675 Ludmilla &  72.09 & S &     6.24 &    0.38 &   0.08 & 76  \\
445 Edna &  87.18 & C &     6.18 &    0.17 &   0.19 & 38  \\
718 Erida &  72.94 & C &     6.13 &    0.79 &   0.23 & 32  \\
162 Laurentia &  99.10 & C &     6.10 &    1.03 &   0.62 & 35  \\
375 Ursula & 182.53 & C &     6.05 &    5.83 &   1.50 & 64  \\
381 Myrrha & 120.58 & C &     5.86 &    0.17 &   0.14 & 49  \\
1001 Gaussia &  74.68 & C &     5.68 &    0.24 &   0.12 & 36  \\
306 Unitas &  46.70 & S &     5.64 &    0.43 &   0.15 & 72  \\
690 Wratislavia & 134.66 & C &     5.63 &    4.23 &   1.38 & 61  \\
1013 Tombecka &  31.92 & S &     5.62 &    0.01 &   0.02 & 26  \\
1694 Kaiser &  29.06 & C &     5.62 &    0.16 &   0.07 & 35  \\
76 Freia & 183.66 & C &     5.61 &    2.84 &   1.24 & 44  \\
55 Pandora &  66.70 & S &     5.57 &    0.46 &   0.04 & 76  \\
43 Ariadne &  65.88 & S &     5.43 &    1.97 &   0.91 & 44  \\
604 Tekmessa &  65.16 & C &     5.43 &    0.45 &   0.30 & 29  \\
253 Mathilde &  58.04 & C &     5.40 &    0.36 &   0.24 & 41  \\
202 Chryseis &  86.16 & S &     5.33 &    1.58 &   0.58 & 39  \\
569 Misa &  72.96 & C &     5.31 &    0.28 &   0.08 & 32  \\
176 Iduna & 121.04 & C &     5.20 &    0.69 &   0.52 & 49  \\
26 Proserpina &  94.80 & S &     5.08 &    3.28 &   1.08 & 52  \\
33 Polyhymnia &  53.68 & S &     5.03 &    0.20 &   0.23 & 38  \\
67 Asia &  58.10 & S &     4.98 &    0.73 &   0.29 & 59  \\
791 Ani & 103.52 & C &     4.97 &    0.38 &   0.06 & 38  \\
980 Anacostia &  86.18 & S &     4.83 &    1.80 &   0.61 & 54  \\
165 Loreley & 154.78 & C &     4.76 &    0.77 &   0.54 & 31  \\
1015 Christa &  96.94 & C &     4.72 &    0.41 &   0.28 & 26  \\
971 Alsatia &  63.76 & C &     4.71 &    0.49 &   0.22 & 30  \\
583 Klotilde &  81.64 & C &     4.65 &    0.30 &   0.02 & 28  \\
952 Caia &  81.60 & C &     4.64 &    1.10 &   0.29 & 31  \\
663 Gerlinde & 100.88 & C &     4.51 &    0.31 &   0.07 & 26  \\
584 Semiramis &  54.00 & S &     4.47 &    0.32 &   0.30 & 42  \\
152 Atala & 122.84 & S &     4.36 &    2.63 &   0.95 & 27  \\
349 Dembowska & 139.78 & S &     4.00 &    3.56 &   1.87 & 41  \\
127 Johanna & 124.55 & S &     3.98 &    0.69 &   0.31 & 26  \\
178 Belisana &  35.82 & S &     3.93 &    0.11 &   0.01 & 25  \\
204 Kallisto &  48.58 & S &     3.91 &    0.49 &   0.15 & 29  \\
65 Cybele & 237.26 & C &     3.86 &    1.83 &   0.57 & 65  \\
126 Velleda &  44.82 & S &     3.75 &    0.30 &   0.12 & 26  \\
328 Gudrun & 122.92 & C &     3.75 &    1.63 &   1.15 & 42  \\
57 Mnemosyne & 112.60 & S &     3.60 &    3.22 &   1.65 & 79  \\
665 Sabine &  51.10 & S &     2.89 &    0.02 &   0.03 & 25  \\

\end{longtable}
}

\end{document}